\documentclass[aip,pof,graphicx,amsmath,amssymb]{revtex4-1}

\draft 

\usepackage{graphicx}
\usepackage{dcolumn}
\usepackage{bm}

\usepackage[utf8]{inputenc}
\usepackage[T1]{fontenc}
\usepackage{mathptmx}
\usepackage{etoolbox}
\usepackage{array}
\usepackage{multirow}
\usepackage{tikz}
\usepackage{subfig}
\usetikzlibrary{positioning}
\usepackage{placeins}
\usepackage{afterpage}
\usepackage{ulem}
\usepackage{listings}
\usepackage{caption,tabularx,booktabs}
\usepackage{siunitx}
\usepackage{lmodern}
\makeatletter
\def\@email#1#2{%
 \endgroup
 \patchcmd{\titleblock@produce}
  {\frontmatter@RRAPformat}
  {\frontmatter@RRAPformat{\produce@RRAP{*#1\href{mailto:#2}{#2}}}\frontmatter@RRAPformat}
  {}{}
}%
\makeatother

\usepackage{url}
\usepackage{xcolor}
\definecolor{newcolor}{rgb}{.8,.349,.1}

\newcommand\bX{\boldsymbol X}

\newcommand\bD{\boldsymbol D}
\newcommand\bA{\boldsymbol A}

\newcommand\bu{\boldsymbol u}

\newcommand\bx{\boldsymbol x}

\usepackage{graphicx}
\usepackage{subcaption}

\newsavebox{\measurebox}
\newcolumntype{L}[1]{>{\raggedright\let\newline\\\arraybackslash\hspace{0pt}}m{#1}}
\newcolumntype{C}[1]{>{\centering\let\newline\\\arraybackslash\hspace{0pt}}m{#1}}
\newcolumntype{R}[1]{>{\raggedleft\let\newline\\\arraybackslash\hspace{0pt}}m{#1}}

\begin{document}

\title{Clustering the Flow: A Data-Driven Framework for Pattern
Discovery in Fluid Dynamics}
%




\author{J.A. Martín}%
\affiliation{E.T.S.I. Aaeron{\'a}utica y del Espacio, Universidad Polit{\'e}cnica de Madrid, Pza. Cardenal Cisneros 3, 28040, Madrid, Spain.}

\author{E. Muñoz}\altaffiliation[Electronic email: ]{eva.munoz.salamanca@ulb.be}
\affiliation{Aero-Thermo-Mechanics (ATM) Department, Université Libre de Bruxelles (ULB), Brussels, 1050 Belgium.
}%
\affiliation{Brussels Institute for Thermal-fluid systems and clean Energy (BRITE), Universit\'{e} Libre de Bruxelles and Vrije Universiteit Brussel, Belgium.}%
\affiliation{E.T.S.I. Aaeron{\'a}utica y del Espacio, Universidad Polit{\'e}cnica de Madrid, Pza. Cardenal Cisneros 3, 28040, Madrid, Spain.}
 

\author{H. Dave}
\affiliation{Aero-Thermo-Mechanics (ATM) Department, Université Libre de Bruxelles (ULB), Brussels, 1050 Belgium.
}%
\affiliation{Brussels Institute for Thermal-fluid systems and clean Energy (BRITE), Universit\'{e} Libre de Bruxelles and Vrije Universiteit Brussel, Belgium.}%
\affiliation{Department of Mechanical Engineering, Indian Institute of Technology (IIT) Jodhpur, Nagaur road, Karwar, Jodhpur 342030, India}



\author{A. Parente}\altaffiliation[Electronic email: ]{alessandro.parente@ulb.be}
\affiliation{Aero-Thermo-Mechanics (ATM) Department, Université Libre de Bruxelles (ULB), Brussels, 1050 Belgium.
}%
\affiliation{Brussels Institute for Thermal-fluid systems and clean Energy (BRITE), Universit\'{e} Libre de Bruxelles and Vrije Universiteit Brussel, Belgium.}%
\affiliation{WEL Research Institute, Avenue Pasteur 6 Wavre, 1300, Belgium.}

\author{S. Le Clainche}\altaffiliation[Electronic email: ]{soledad.leclainche@upm.es}
\affiliation{E.T.S.I. Aaeron{\'a}utica y del Espacio, Universidad Polit{\'e}cnica de Madrid, Pza. Cardenal Cisneros 3, 28040, Madrid, Spain.}

\date{\today}

\begin{abstract}
Clustering techniques offer a powerful framework for analyzing complex flow dynamics and reducing computational costs in large-scale simulations. In this work, we propose a novel clustering-based approach using Vector Quantization Principal Component Analysis (VQPCA) to identify structural sensitivity zones, namely the regions where the fluid flow is more receptive to changes. To the authors’ knowledge, this is the first application of VQPCA to a fluid dynamics problem for the identification of flow patterns and dynamically relevant regions. As a fully data-driven technique, it does not rely on adjoint methods; therefore, this approach has the advantage of having low computational cost, since it depends exclusively on data from the direct problem. The VQPCA technique demonstrates its ability to extract dominant flow features by clustering the flow field into regions characterized by their intrinsic dynamics. To assess the validity of this method, it is used to investigate the wake behind a circular cylinder, revealing similarities to previously established structural sensitivity regions. The robustness of the approach is further assessed through validation and calibration in different operating conditions in this flow scenario. As an extension of the analysis, we address the complex dynamics of two planar synthetic jets, where the clustering insights can lead to develop flow control strategies. These results highlight the potential of clustering-based methods as practical and effective tools to analyze and optimize fluid flows.
\end{abstract}


\maketitle 

\section{Introduction}
Understanding the intrinsic dynamics of complex fluid flows is crucial for diverse industrial fields, such as aerospace and wind engineering, oil and gas industry, or medicine. However, analyzing these flows is challenging, 
as they involve a broad range of scales\cite{Peters2009Multiscale} (in both space and time), nonlinear interactions\cite{Holmes2012Turbulence}, and experimental\cite{Orlu2017Reynolds} and computationally\cite{Jimenez2025Chaos} expensive simulations. An approach to address these drawbacks is modal decomposition techniques, which extract dominant flow structures and provide a reduced-order representation of the system\cite{Tairaetal2017}. These methods have been proven to be very effective in capturing the essential physics of the flow while maintaining a moderate computational cost.

By identifying the dominant modes that govern the main dynamics of the flow, modal decomposition can be integrated with other strategies for specific applications such as instability analysis, flow control, or optimization. A complementary approach involves the identification of structural sensitivity zones, which highlight regions where small perturbations can have a significant global impact. As originally described by Giannetti and Luchini\cite{Giannetti2007}, structural sensitivity provides a powerful framework to detect regions of high receptivity to perturbations, which makes it particularly useful for guiding flow modifications.

A key challenge in computing structural sensitivity is its high computational cost, as traditional methods require solving both direct and adjoint problems\cite{Giannetti2007}. To address this limitation, clustering techniques provide an efficient, data-driven alternative to identify dynamically relevant flow regions directly from simulation data.

In this work, we adopt the term structural sensitivity zones to describe flow regions where small structural perturbations produce significant dynamic effects, and therefore small changes are most likely to influence the global dynamics.  
Here, such zones are identified through a data-driven approach based solely on direct flow simulations. This broadens the concept by allowing its application to unsteady and nonlinear regimes where adjoint solutions are difficult or expensive to compute.


In this study, we explore the use of Vector Quantization Principal Component Analysis (VQPCA) to systematically identify structural sensitivity zones, reducing computational costs while maintaining accuracy. This clustering-based approach not only enables a deeper understanding of flow organization but also offers valuable insights for applications such as reduced-order modeling and flow control. To the authors’ knowledge, this is the first application of VQPCA to a fluid dynamics problem for the identification of flow patterns and dynamically relevant regions.

From a conventional perspective, the computation of structural sensitivity maps typically involves combining direct and adjoint eigenfunctions. Luchini, Giannetti, and Pralits~ \cite{LuchiniGiannettiPralits2008} developed a method based on this approach, which, despite its accuracy, can be computationally expensive, particularly for realistic three-dimensional flows. 
Recently, a novel approach~\cite{CorrochanoLeClainche2022} has been proposed to compute sensitivity maps at a reduced computational cost, since it requires only the computation of the direct problem. However, neither of these approaches can be extended to unsteady problems as they consider structural variations in the mean flow.

Even when sensitivity maps are available, identifying key regions for further analysis remains a challenge, particularly in complex flow environments. This issue is especially relevant when external modifications are introduced in these sensitive areas to influence global flow instabilities, as in the case of flow control. Techniques in this field are generally classified as active (e.g., oscillations and jets) or passive (e.g., cavities and vortex generators~\cite{WangFeng2018}). The process of identifying the most effective approach typically follows two main steps: first, selecting the appropriate technique, and second, optimizing the control laws through numerical procedures. For nonlinear problems, genetic algorithms~\cite{DuriezBruntonNoack2017} are commonly used due to their ability to handle complex search spaces; however, they require hundreds of thousands of evaluations to achieve convergence. To make this computationally feasible, surrogated models are often used to approximate the response of the system, although the overall cost can still be significant~\cite{PehlivanogluYagiz2012}. In this context, clustering techniques provide a powerful tool for analyzing complex flow dynamics by identifying dynamically relevant regions. As a direct application, they can enhance the control optimization step by reducing the search domain. 


The primary objective of this work is to investigate a clustering technique as a tool to locate regions of maximum structural sensitivity in both steady and unsteady problems, substantially reducing computational cost. Clustering allows partitioning the flow field into dynamically relevant regions and provides a data-driven approach to identify areas where small perturbations can have a significant impact. 


The proposed clustering method, Vector Quantization Principal Component Analysis (VQPCA)~\citep{Parenteetal2011}, partitions the flow domain into distinct clusters. This study explores the potential connection between these clusters and structural sensitivity regions. VQPCA is based on Principal Component Analysis (PCA), a well-established dimensionality reduction technique. Unlike traditional PCA, which operates on the entire dataset, VQPCA applies PCA locally within clusters, which are strategically chosen by the algorithm to optimize field reconstruction. While VQPCA has been widely used in combustion datasets to build reduced-order models~\citep{DAlessioAttilietal2021, AmaduzziDAlessioetal2024} and feature extraction~\citep{ZdybalDAlessioAttilietal2023}, recent work by Muñoz et al.~\cite{MunozDaveetal2023} has demonstrated its potential in non-reactive flows, where it effectively identifies and categorizes regions based on their dynamics.


The approach is first applied to the flow past a cylinder, a case widely studied with well-established structural sensitivity regions. Given the extensive research on the inherent instabilities in this flow~\cite{Strykowski1990}, it is the best candidate as a test case to calibrate the algorithm and validate our hypothesis. To further evaluate the robustness of the method, we analyze three datasets covering both two- and three-dimensional flows while varying the Reynolds number and spatial sampling.\\
After calibration, the algorithm is applied to the flow generated by two synthetic jets. These devices play a crucial role in various industrial applications, from improving fluid mixing~\citep{Wang2001}, to optimizing heat transfer~\citep{Pavlova2006} and even as alternative propulsion systems, inspired by the swimming motion of marine animals such as squids and jellyfish~\citep{Demont1988}.\\
Subsequently, the identified structural sensitivity zones, determined through VQPCA, are analyzed to explore their potential for modulating flow instabilities. Although these regions provide valuable information about the dynamics of the fluid, flow control naturally emerges as a potential application of this technique. To further illustrate this, two flow perturbation methods are examined: (i) active perturbation via point force application and (ii) passive perturbation through the introduction of flow disruptors.

This study is organized as follows. Section~\ref{sec:Methodology} presents the methodology used, covering the clustering algorithm. The calibration and validation of this technique, performed on the flow around a circular cylinder, are detailed in Section~\ref{sec:Results_Cyl}. Section~\ref{sec:Results_Jets} explores the application of the algorithm to synthetic jets, identifying structural sensitivity zones where a potential control strategy could be applied. Section~\ref{sec:Conc} summarizes the key conclusions.

\section{Methodology}\label{sec:Methodology}

The governing equations describing an incompressible flow are the continuity and Navier-Stokes (NS) equations. They are written in a non-dimensional form as:
\begin{align}
   \nabla \cdot \bu &= 0, \label{eq:cont} \\
  \frac{\partial \bu}{\partial t} + \left( \bu \cdot \nabla \right) \bu &= -\nabla p + \frac{1}{Re} \Delta \bu, \label{eq:NS}
\end{align}
where $\bu$ and $p$ are the nondimensional velocity vector and pressure respectively and Re=$UL/\nu$ is the Reynolds number, being $U$ and $L$ the characteristic flow velocity and length, and $\nu$ the kinematic viscosity of the fluid. These equations are nondimensionalized with the units for length and time, $L$ and $L/U$, respectively, as usual.

\subsection{Structural sensitivity}\label{subsec:struct_sen}

In the field of fluid dynamics, Giannetti and Luchini~\cite{Giannetti2007} introduced the concepts of structural sensitivity and wavemakers. These concepts aim to identify flow regions where structural modifications in the stability problem lead to the most pronounced drift of the dominant linear stability eigenmode. A Reynolds decomposition is applied to the flow, which is expressed as a combination of the base flow or expectation ($\overline{\bu}, \overline{p}$) and an infinitesimal unsteady perturbation ($\bu', p'$), represented as
$$\bu = \overline{\bu}+ \bu' \ \ \text{and} \ \ p=\overline{p}+p'.$$
Substituting this decomposition into Equation \ref{eq:NS}, neglecting higher-order nonlinear terms in $\bu'$, and introducing a forcing term, denoted as $\boldsymbol{f}$, leads to:
\begin{equation}
   \frac{\partial \bu'}{\partial t} + \left( \overline{\bu} \cdot \nabla \right) \bu' + \left( \bu' \cdot \nabla \right) \overline{\bu} =  -\nabla p' + \frac{1}{Re} \Delta \bu' + \boldsymbol{f}. \label{eq:NS_force}
\end{equation}

For structural modifications, the forcing term is modeled to be proportional to the eigenmodal solution of the flow fluctuations. In linear flows, maximum sensitivity (maximum eigenvalue variation) is defined as the product of the magnitude of direct and adjoint velocity fields~\citep{Marquet2008}.  Marquet et al~\cite{Marquet2008} compared the impact of both base flow modifications and the application of a steady force. 
Following this approach, the present study examines both strategies within the identified sensitivity zones, considering the use of point forces and base flow modifications through the introduction of disruptors. In this context, "disruptors" refer to square elements strategically placed within the domain, subject to specific wall conditions.

However, identification of structural sensitivity zones using conventional methods is highly computationally demanding, as it requires solving both the direct and adjoint Navier-Stokes equations. This challenge becomes even more prohibitive for unsteady applications~\cite{Rubinoetal2020}. To overcome this limitation, we propose an alternative approach based on VQPCA, which identifies potential structural sensitivity zones while only requiring the solution of the direct problem. This method not only extends the applicability of sensitivity analysis to unsteady flows, but also significantly reduces computational costs~\cite{DaveSwaminathanParente2022}.


\subsection{Clustering method: Vector Quantization Principal Component Analysis (VQPCA)} \label{sec:Meth_VQPCA}

Vector Quantization via Principal Component Analysis (VQPCA) is a local PCA technique introduced by Kambhatla and Leen~\cite{KambhatlaLeen1997} to mitigate the reconstruction errors associated with the linearity of traditional PCA. By applying PCA locally, VQPCA constructs a more compact low-dimensional manifold that better captures the intrinsic structure of the data. This section provides an overview of the algorithm for this study. For more detailed information, please refer to Ref.~\cite{Parenteetal2011}.


In essence, VQPCA partitions the input dataset into distinct clusters of similar data points, acting as an unsupervised clustering algorithm. The input data consists of the flow dataset to be segmented, along with two crucial hyperparameters: the number of principal components (PCs) retained in each cluster, denoted by $q$ and the desired number of clusters, $k$. The input dataset is structured as a matrix $\mathbf{X} \in \mathbb{R}^{n \times p}$, where $n$ represents the number of statistical observations to cluster and $p$ is the number of original variables.


Before applying machine learning algorithms to multivariate data, it is common practice to preprocess the dataset through centering and scaling. These operations are mathematically expressed as: $\bX'=(\bX - \overline{\bX})\bD^{-1}$, where $\overline{\bX}$ and $\bD$ represent the centering and scaling matrices, respectively. From this point on, the preprocessed matrix is referred to as $\bX$.  The centering step subtracts the mean value of each variable, while scaling normalizes each variable by the square root of its variance. For a more detailed explanation of these operations, refer to Ref.~\cite{ParenteSutherland2013}.

\begin{figure}[h]
  \centering
  \includegraphics[width=0.8\textwidth]{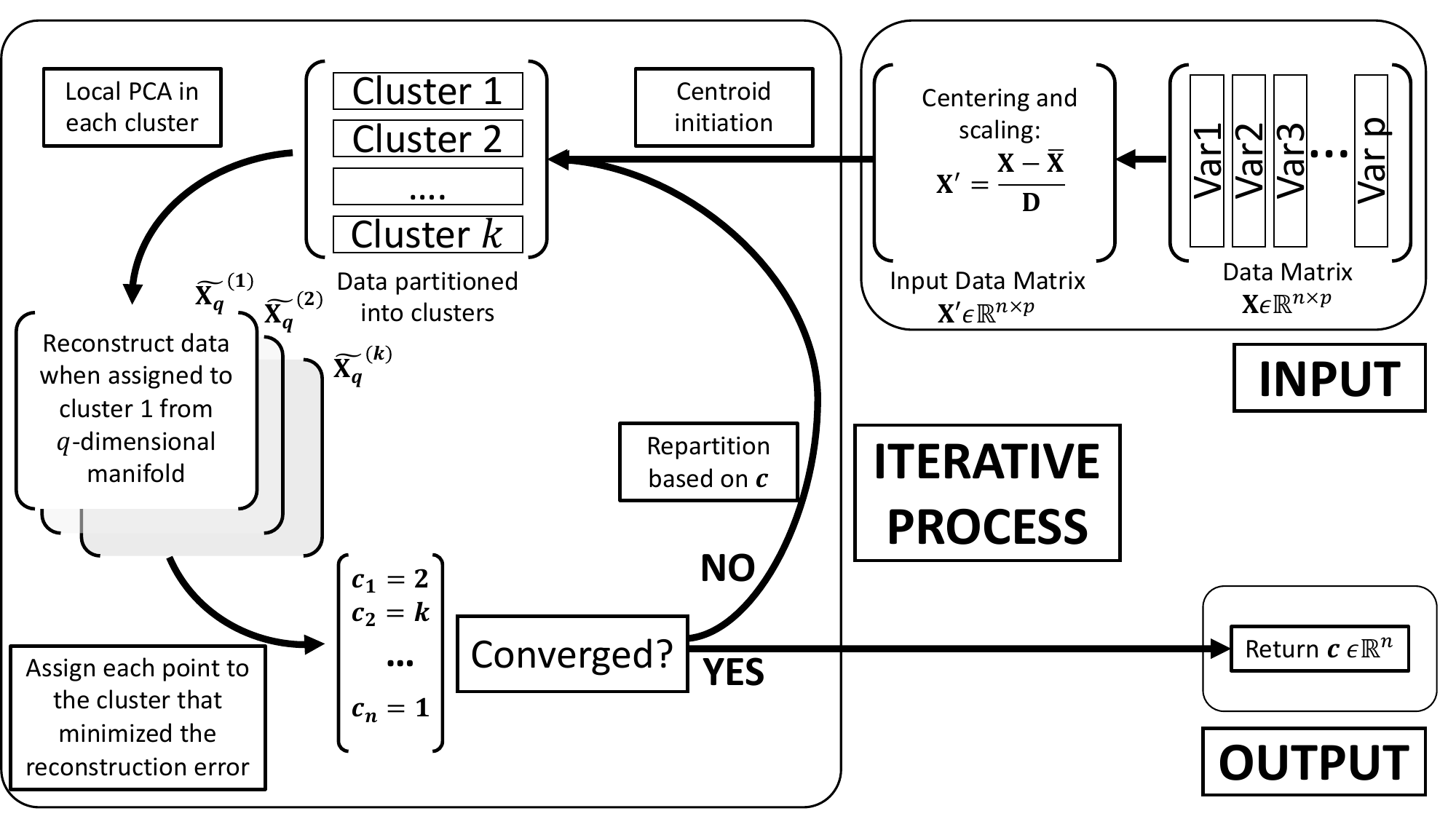}
  \caption{Scheme of VQPCA algorithm, adapted from Ref.~\citep{DaveSwaminathanParente2022}}
  \label{fig:Meth_VQPCA}
\end{figure}

Figure~\ref{fig:Meth_VQPCA} illustrates the steps undertaken by the algorithm. The process begins with the initialization of the cluster centroids, which is selected to be a converged k-means solution~\citep{MacQueen1967} in order to accelerate convergence~\citep{DAlessioParenteetal2020}. For each cluster, the local eigenvector matrix, $\bA^{(k)}$, is computed, and a low-dimensional space, denoted as $\bA_q^{(k)}$, is defined by selecting the first \(q\) principal components (\(q < p\)). The reconstructed data, $\widetilde{\bX}_q^{(k)}$, are then obtained using the reduced basis. The reconstruction error at each grid point $i$ is calculated as $\epsilon_i^{(k)} = | \bx_i - \widetilde{\bx}_{i,q}^{(k)} |$. Subsequently, each data point is assigned to the cluster with the minimum reconstruction error, and the cluster centroids are updated. This process is repeated until the variation of the centroids falls below a predefined tolerance~\citep{DaveSwaminathanParente2022}.

The application of VQPCA to analyze complex fluid flows requires a specific arrangement of the data matrix, including compression along the temporal dimension. This approach enables the construction of a low-dimensional manifold that captures dominant flow patterns~\citep{MunozDaveetal2023}, allowing the domain to be partitioned into regions with similar dynamical behavior. Although the primary aim of this clustering technique is to reveal the underlying flow organization, one of its practical applications lies in flow control. Control strategies often rely on the introduction of perturbations (Eq.~\ref{eq:NS_force}) to modify the dynamics of the system, typically to suppress or enhance instabilities. Identifying the regions where such perturbations have the greatest impact is essential. Since VQPCA partitions the domain based on dominant flow dynamics, which are intrinsically related to flow instabilities, it can highlight regions of heightened sensitivity. 
The underlying hypothesis and its calibration are assessed using the canonical case of a flow past a cylinder (Sect.~\ref{sec:Results_Cyl}).

\section{Application on the wake of a circular cylinder}\label{sec:Results_Cyl}

\subsection{Instabilities on the wake of a circular cylinder}\label{sec:Cyl_Instabilities}

The flow past a circular cylinder is a classic and well-understood benchmark problem in fluid dynamics. Despite its apparent simplicity, a range of flow regimes emerges as the Reynolds number increases. The Reynolds number is defined as $\mathrm{Re} = \frac{U D}{\nu}$, where $U$ is the incoming free-stream velocity, $D$ is the cylinder diameter, and $\nu$ is the kinematic viscosity of the fluid. At low Reynolds numbers ($\mathrm{Re} < 40$), the flow remains steady and symmetric with respect to the mid-$Y$ and mid-$Z$ planes, and it is essentially two-dimensional. As $\mathrm{Re}$ increases to around 46, the first Hopf bifurcation occurs~\citep{Jackson1987}, resulting in an unsteady and periodic vortex shedding known as the von Kármán vortex street, though the flow remains two-dimensional. At approximately $\mathrm{Re} \approx 189$, a secondary bifurcation triggers the onset of three-dimensionality, which depends on the spanwise length of the cylinder, defined as $L_z = 2\pi/\beta$, where $\beta$ is the spanwise wavenumber~\citep{BarkleyHenderson1996}.


The sensitivity of vortex shedding to small external perturbations~\citep{Kovasznay1949} and geometrical modifications~\citep{Stansby1974} at low Reynolds numbers has been well-established in previous studies. Strykowski and Sreenivasan~\cite{Strykowski1990} identified specific regions in the wake where local alterations exert the greatest influence on the global flow. Their analysis, involving structural perturbations introduced through smaller cylinders, aimed to control the onset of flow instability and was supported by both experimental and computational investigations. Later, Giannetti and Luchini~\cite{Giannetti2007} introduced a structural sensitivity analysis based on linearization around the base flow, defining these critical areas as structural sensitivity zones. In particular, the regions identified in the two studies do not overlap completely. This discrepancy arises because the former captures the effect of perturbations on both the base flow and its fluctuations, while the latter considers only linear perturbations around a steady base flow.


In this section, we apply VQPCA to the cylinder datasets presented in Sec.~\ref{sec:Cyl_Datasets} to determine whether the same structural sensitivity zones can be recovered. These results serve to validate the proposed clustering-based approach and to calibrate the hyperparameter selection for future applications.


\subsection{Datasets}\label{sec:Cyl_Datasets}

This study employs three distinct datasets representing the wake behind a circular cylinder at different Reynolds numbers. Two datasets correspond to two-dimensional simulations at Reynolds numbers of 60 and 100, while the third represents a three-dimensional flow at a Reynolds number of 280, with a spanwise domain length of 6.99. For clarity, these cases are referred to as \textit{2D60}, \textit{2D100}, and \textit{3D280} throughout the text.


Numerical simulations were conducted using the open-source spectral element solver \citep{nek5000}, which solves the incompressible continuity and Navier–Stokes equations in non-dimensional form (Eqs.\ref{eq:cont} and~\ref{eq:NS}). The computational domain extends from $x \in [-15, 50]D$ in the streamwise direction and $y \in [-15, 15]D$ in the normal direction. Boundary conditions were implemented following the setup described in Ref.~\cite{BarkleyHenderson1996}. At the inlet and on the upper and lower boundaries, Dirichlet conditions were applied for the velocity field ($u_x = U = 1$, $u_y = 0$, and $u_z = 0$ in the three-dimensional case), while Neumann conditions were used for pressure. Standard outflow conditions were imposed at the outlet, consisting of Neumann conditions for velocity and Dirichlet for pressure. No-slip conditions were enforced on the cylinder surface. The two-dimensional domain consists of 600 rectangular macro-elements, each discretized with $p+1$ Gauss–Lobatto–Legendre points, where $p = 9$ is the polynomial order. For the three-dimensional simulation, the spanwise direction is extruded using 64 equispaced planes over a wavelength of $L_z = 6.99D$, with periodic boundary conditions imposed at the spanwise ends. Additional details about the simulation setup and parameters can be found in Refs.~\cite{HODMDbook, LeClainchePerezVega2018}.


Although the three datasets share the same computational domain in the $x$-$y$ plane, they differ in their temporal and spatial sampling. Details of these sampling schemes are provided in Table~\ref{tab:Cyl_grid}. The temporal sampling is characterized by the number of equispaced snapshots, $N_t$, collected during the transient or saturated regime, after the simulations have converged in time. The corresponding time intervals, $\Delta t$, are also presented for each of the three test cases. The temporal sampling of the three-dimensional simulation is limited to transient data due to an instability in the spanwise direction, which evolves over time as the flow saturates in the other spatial directions~\citep{LeClainchePerezVega2018}.

For spatial sampling, the number of grid points ($N_x$, $N_y$, $N_z$) in the streamwise, normal and spanwise directions is specified, along with their respective locations ($x$, $y$, $z$). The center of the cylinder is located at the origin of the coordinate system ($x = 0$, $y = 0$), and in the three-dimensional case, its length is given by $L_z$.


\begin{table}[h!]
  \begin{tabular}{|c|c| ccc|}
  \hline
     Sampling  				& 	Simulation         		& \textit{2D60}  			& \textit{2D100} 	& \textit{3D280} 	\\ \hline \hline
    \multirow{3}{*}{ Temporal}  	&  Time interval $\Delta t$ 	&   0.2 $D/U$   			&   0.2 $D/U$		&  1.0 $D/U$       	 \\ 
   					 	&  $N_t$ 					&   151    				&   151      		&  600      			 \\
     						&  Regimen       			& Saturated   			&  Saturated    		& Transitory    		 \\ \hline \hline
     \multirow{2}{*}{ Streamwise} &Location  				& $x \in [-9.2, 26.5]D$ 	&$x \in [-1, 8]D$ 	& $x \in [0, 10]D$   	\\
     						& $N_x$        				&    1000        			&  449         		&    200        		\\ \hline
    \multirow{2}{*}{Normal}  	& Location    				& $y \in [-16.8, 18.9]D$ 	&$y \in [-2, 2]D$ 	&  $y \in [-2, 2]D$  	\\
    					 	& $N_y$        				&    1000       			&  199             		&    40        		 \\ \hline
   \multirow{2}{*}{Spanwise}   	& Location   				&    -          			&  -         			& $z \in [0, 6.99]D$	 \\
    					    	& $N_z$        				&    -         				&  -         			&    64             		\\ \hline
  \end{tabular}	
  \caption{Summary of the temporal and spatial sampling of each dataset. Temporal sampling: number of snapshots collected, $N_t$, their time interval, $\Delta t$ and the regimen in which they are collected: saturated or transitory. Spatial sampling: number of points, $N_x$, $N_y$ and $N_z$, and their location in streamwise, normal, and spanwise directions. }
  \label{tab:Cyl_grid}
\end{table}

\begin{figure}[htb]
\begin{tabular}{cc}
\includegraphics[width=0.35\textwidth]{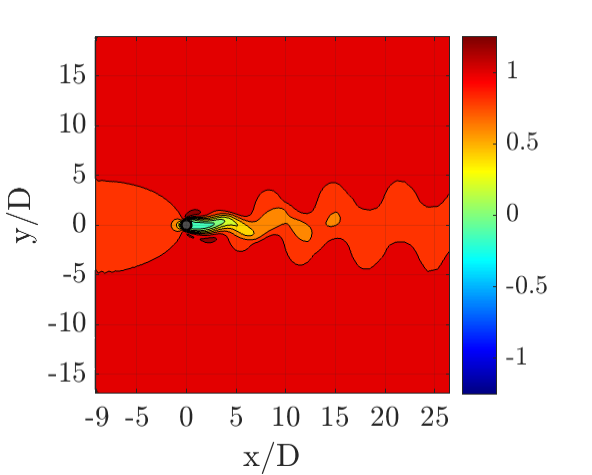}\put(-160,-10){\footnotesize (a) Streamwise component, $u_x$,  of \textit{2D60}}&
~~~~~~\includegraphics[width=0.35\textwidth]{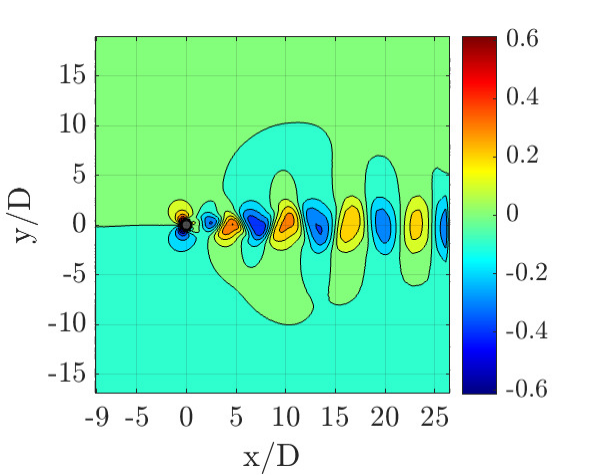}\put(-160,-10){\footnotesize (b) Normal component, $u_y$,  of \textit{2D60}}\\
&\\
\includegraphics[trim ={0.0cm 1.0cm 0.2cm 2.2cm},clip, width=0.45\textwidth]{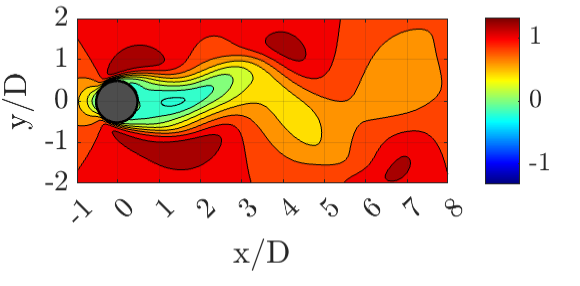}\put(-185,-7){\footnotesize (c) Streamwise component, $u_x$, of \textit{2D100}}&
~~~~~~\includegraphics[trim ={0.0cm 1.0cm 0.2cm 2.2cm},clip, width=0.45\textwidth]{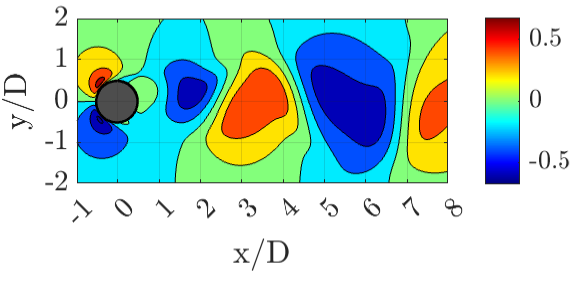}\put(-185,-7){\footnotesize (d) Normal component, $u_y$, of \textit{2D100}}
\end{tabular}
  \caption{Contour plots of the velocity components of \textit{2D60} and \textit{2D100} in their sampling domain.}
 \label{fig:2D60y100}
\end{figure}

Figure~\ref{fig:2D60y100} shows contour plots of (a) the streamwise velocity component, $u_x$, and (b) the normal velocity component, $u_y$, for the \textit{2D60} and \textit{2D100} datasets. As expected, the first instability is evident in both cases, marked by the vortex shedding downstream of the cylinder, consistent with Reynolds numbers exceeding $46$. Additionally, these plots highlight the differences in spatial sampling, displaying the complete computational domain used for the data analysis, as detailed in Table~\ref{tab:Cyl_grid}.


\begin{figure}[htb]
\begin{tabular}{cc}
\includegraphics[width=0.45\textwidth]{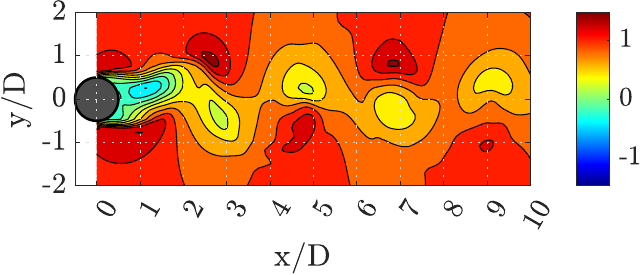}\put(-210,-10){\footnotesize (a) Contours of the streamwise component of the velocity,}\put(-210,-22){\footnotesize $u_x$ in the middle $x$-$y$ plane}&
~~~~~~\includegraphics[trim ={4.0cm 2.0cm 7.0cm 4.0cm},clip, width=0.45\textwidth]{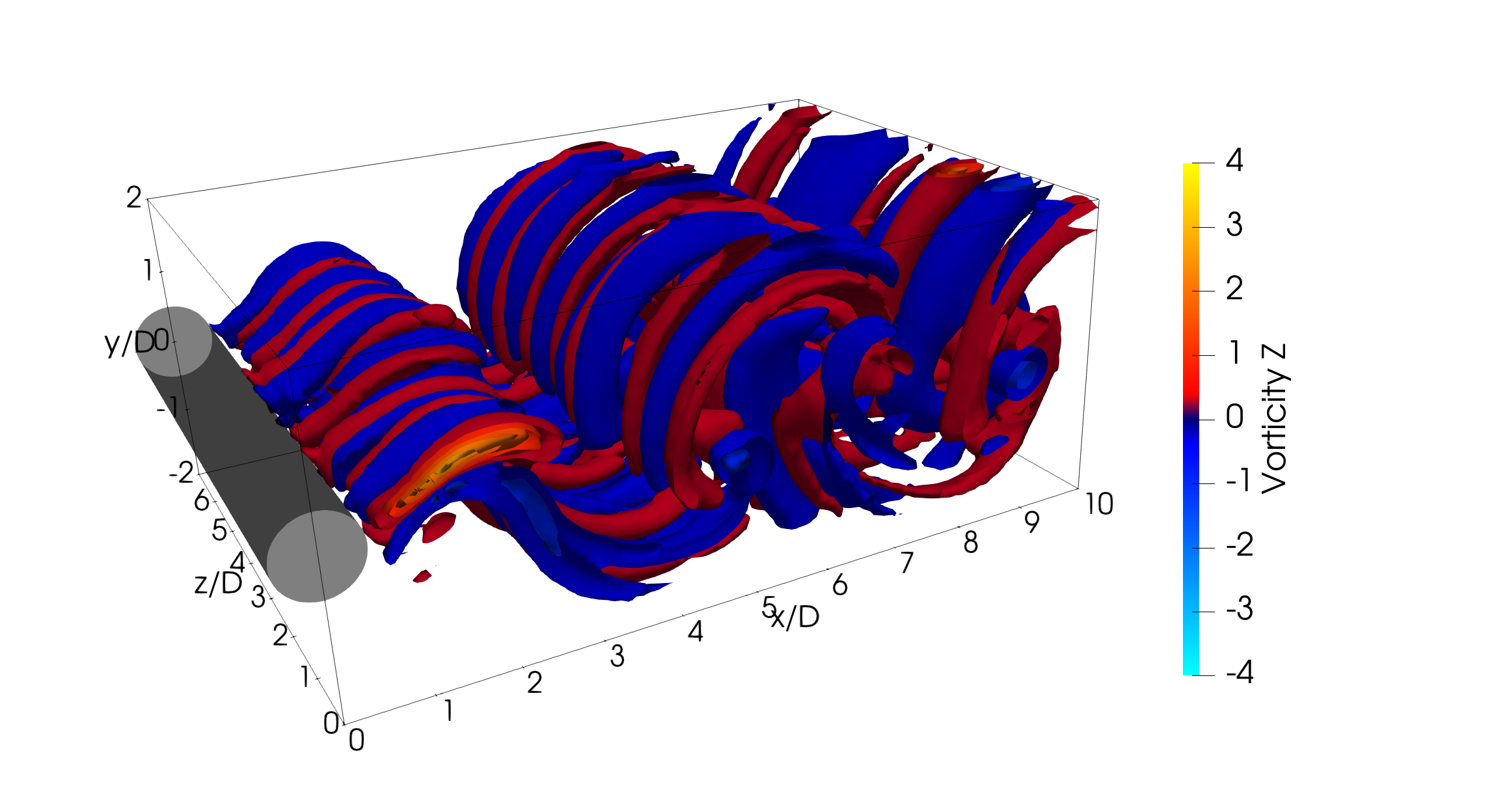}\put(-180,-10){\footnotesize (b) Isosurface plot of the vorticity, $\omega_z$}
\end{tabular}
  \caption{(a) Streamwise component of the velocity $u_x$ and (b) vorticity $\omega_z$ of \textit{3D280}}
 \label{fig:3D280}
\end{figure}

Figure~\ref{fig:3D280} displays the streamwise velocity component in the middle $x$-$y$ plane (left panel) and the three-dimensional spanwise vorticity, defined as $\omega_z = \frac{\partial u_y}{\partial x} - \frac{\partial u_x}{\partial y}$ (right panel), for the \textit{3D280} case. As shown, this case exhibits structures not only in the $x$-$y$ plane but also along the spanwise direction, indicating that the second bifurcation has occurred. This is consistent with the Reynolds number exceeding $189$ and the spanwise length of the cylinder being sufficient for the instability to develop.

The variety of datasets allows for a comprehensive evaluation of the algorithm's performance under (i) varying operating conditions and (ii) different spatial dimensionalities. This is achieved by analyzing two- and three-dimensional datasets that capture diverse spatial domains at different Reynolds numbers.



\subsection{Data preparation}\label{sec:Cyl_DataPrep}
The datasets have been structured to ensure consistency in the application of the clustering technique. Each dataset is organized into a matrix $\bX \in \mathbb{R}^{n \times p}$, where $n$ represents the number of statistical observations (snapshots), and $p$ is the number of variables representing the dimension to be reduced. For convenience, the total number of grid points in space is defined as $N_{xyz} = N_x N_y N_z$, with $N_z = 1$ for the two-dimensional cases. The number of selected fluid variables (e.g., velocity components and pressure) is denoted by $N_v$, taking values $2$ and $3$ for 2D and 3D cases respectively (representing $u_x$, $u_y$ and $u_z$), and the total number of snapshots included in the analysis is $n_t$. Therefore, each snapshot in the dataset corresponds to a concatenated vector of size $p = N_{xyz} \cdot N_v$.

\begin{figure}[h!]
 \centering
\includegraphics[width=0.35\textwidth]{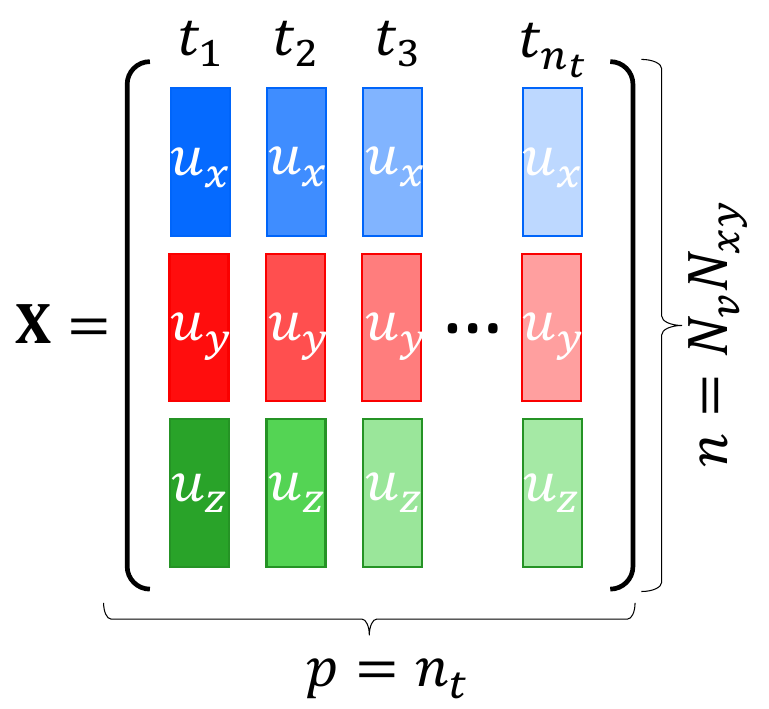}\label{subfig:scheme_Xtime}
 \caption{Scheme of the matrix $\bX$}
 \label{fig:scheme_X}
\end{figure}

Following Muñoz et al.~\cite{MunozDaveetal2023}, the data are organized in matrices with dimensions $n=N_{xyz} N_v$  and $p=n_t$. Figure~\ref{fig:scheme_X} schematizes the construction of the matrices.

\subsection{Clustering results}\label{sec:Cyl_clus}

This section presents the results of the VQPCA implementation in the cylinder datasets (Sec.~\ref{sec:Cyl_Datasets}), described in Sec.~\ref{sec:Cyl_DataPrep}, with the goal of identifying structural sensitivity zones.

To begin the analysis, the first step involves calibrating the hyperparameters of the algorithm. To capture the most energetic direction in time, the number of principal components retained is fixed at $q=1$. The number of selected snapshots, $n_t$, is chosen to cover a full number of oscillation cycles in periodic flows, while being kept as low as possible to reduce computational cost. The number of clusters, $k$, is determined by minimizing the Davies-Bouldin Index (DBI)~\citep{DaviesBouldin1979}, which quantifies cluster quality by balancing intra-cluster similarity and inter-cluster separation. Section~\ref{subsec:hyperp} details the hyperparameter calibration using the \textit{2D100} dataset as a reference case. This dataset was selected due to its lower spatial dimensionality, which enables faster evaluations during the calibration phase. Once the optimal parameters are identified, the algorithm is systematically applied to the remaining datasets. The results are then analyzed and compared to validate the methodology, as discussed in Sec.~\ref{subsec:cyl_comp}.


The computational cost of the algorithm depends on the selected hyperparameters, as the process is iterative. For reference, the algorithm took $42$ seconds to converge for the \textit{2D100} dataset ($N_{xy}=89351$)  using $k=3$ clusters, $q=1$ retained principal component per cluster and $n_t=30$ snapshots. This result was measured on a single core of an Intel Xeon E5-2603 processor.

\subsubsection{Hyperparameters calibration}\label{subsec:hyperp}

The first hyperparameter to be defined is the number of selected snapshots, denoted as $n_t$. Following the aforementioned guidelines, $n_t$ is set to $30$ for both \textit{2D60} and \textit{2D100}, capturing one full vortex shedding cycle during the quasi-steady regime.  This is sufficient due to the two-dimensional nature of the flow and the regularity of the shedding pattern, which is well represented within a single period. For \textit{3D280}, however, a larger value of $n_t = 40$ is chosen to represent $10$ cycles within the saturated state of the flow, sampled starting from the $100^{\mathrm{th}}$ cycle. In the three-dimensional case, extended sampling ensures that the more complex and potentially asymmetric three-dimensional structures are adequately captured. This information is summarized in Table~\ref{tab:nt_selected}.


\begin{table}[h]
  \centering
  \begin{tabular}{|l|ccc|}
  \hline
    Simulation & \textit{2D60} & \textit{2D100} & \textit{3D280}   \\ \hline
    Snapshots selected $n_t$ & 30 & 30 & 40  \\
    Wave cycles on the interval & 1 & 1 & 10 \\ \hline
  \end{tabular}
  \caption{Summary of the number of snapshots selected, $n_t$, and corresponding wave cycles in each case.}
  \label{tab:nt_selected}
\end{table}

\begin{figure}[h!]
  \centering
  \includegraphics[width=\textwidth]{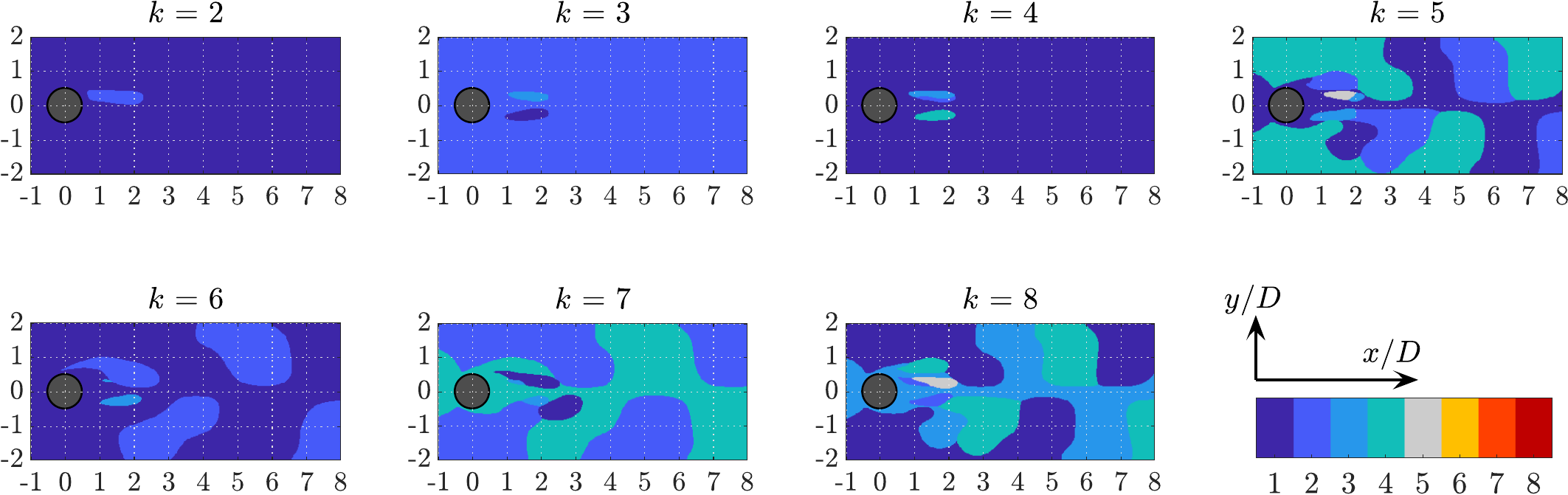}
  \caption{Clustering contours corresponding to the streamwise velocity component, $u_x$, of the reference case \textit{2D100}. Each subfigure corresponds to a number of clusters, $k \in [2,8]$.}
  \label{fig:VQPCA_k_u}
\end{figure}

\begin{figure}[h!]
  \centering
  \includegraphics[width=\textwidth]{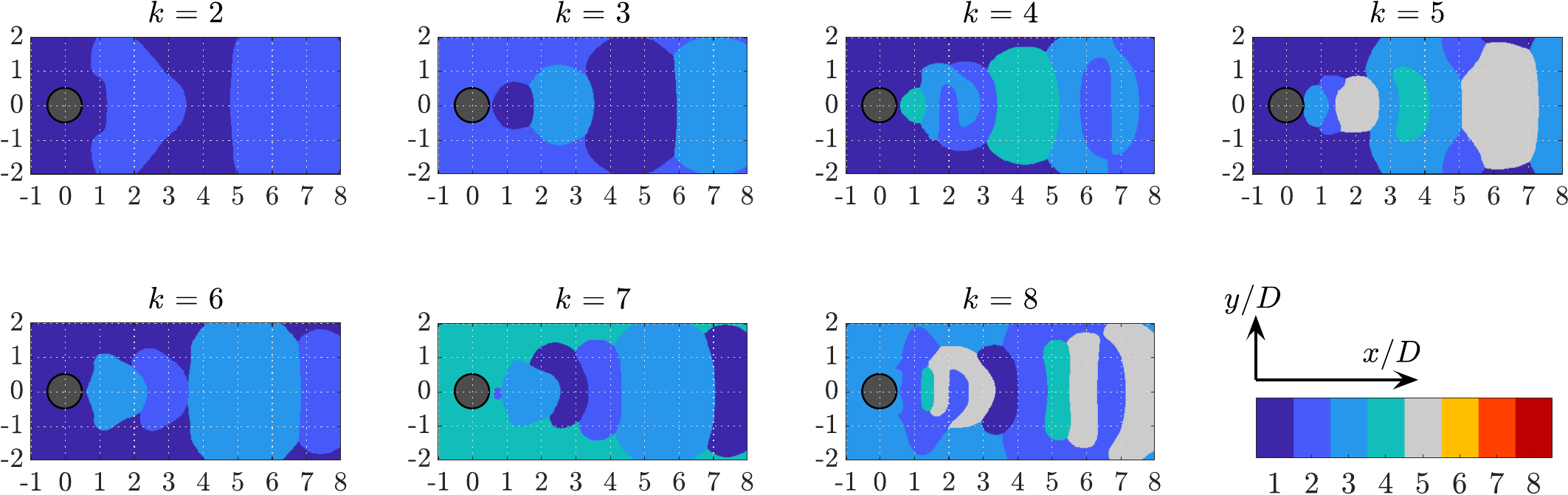}
  \caption{Same as Fig.~\ref{fig:VQPCA_k_u} on the normal component of the velocity, $u_y$.}
  \label{fig:VQPCA_k_v}
\end{figure}

The second and final hyperparameter to be selected is the number of clusters, $k$. To investigate its effect, Figures~\ref{fig:VQPCA_k_u} and \ref{fig:VQPCA_k_v} show the clustering indices obtained for the streamwise and normal velocity components ($u_x$ and $u_y$, respectively), in a range of $k \in [2,8]$. As $k$ increases, the clustering algorithm refines the domain segmentation by either splitting existing regions or introducing entirely new ones. This behavior reflects the ability of the algorithm to identify increasingly localized dynamics within the flow. In particular, while smaller values of $k$ capture the global structure of the flow, larger values provide a more granular view that may better align with local instabilities or flow features relevant to control.


\begin{figure}[htb!]
  \centering
  \includegraphics[width=0.5\textwidth]{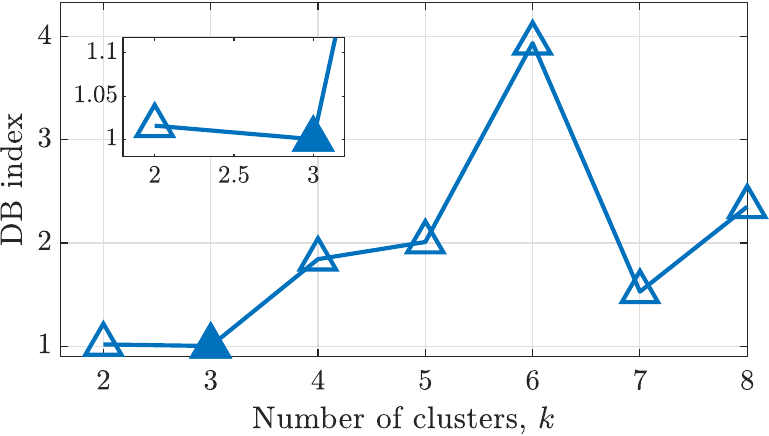}
  \caption{Davies Bouldin (DB) index for the clustering results of \textit{2D100} with $k\in[2,8]$. Optimal solution represented with a filled symbol}
  \label{fig:DBI_cyl}
\end{figure}

As the number of clusters increases, smaller scales are captured. However, the objective is not simply to solve smaller structures, but to identify a partitioning in which each cluster exhibits strong internal coherence in its reduced-dimensional representation while remaining as distinct as possible from the others. In this context, an ideal clustering solution maximizes intra-cluster similarity and inter-cluster dissimilarity, leading to high orthogonality between clusters. To determine the optimal number of clusters, we employ the Davies-Bouldin (DB) index, a commonly used metric based on Euclidean distances that quantifies this trade-off. A lower DB index indicates better clustering performance. Figure~\ref{fig:DBI_cyl} shows the DBI values obtained for the \textit{2D100} dataset as the number of clusters varies from $k=2$ to $k=8$. According to this criterion, the optimal clustering corresponds to $k=3$, which is marked in the figure with a filled symbol.


The clustering contours for the streamwise velocity component $u_x$, shown in Fig.~\ref{fig:VQPCA_k_u} for the optimal configuration ($k=3$), reveal two lobe-shaped regions, while the third cluster spans the remainder of the domain. These lobe structures closely resemble those reported in structural sensitivity analyses of the circular cylinder wake, such as Fig.20 in Ref.\cite{Strykowski1990} and Fig.17 in Ref.\cite{Giannetti2007}, although obtained using different methodologies and Reynolds numbers. As discussed in the Introduction, structural sensitivity zones highlight flow regions that are particularly receptive to perturbations. Introducing minor alterations within these zones can influence the onset of instabilities, either delaying or promoting them, which facilitates the application of flow control strategies.

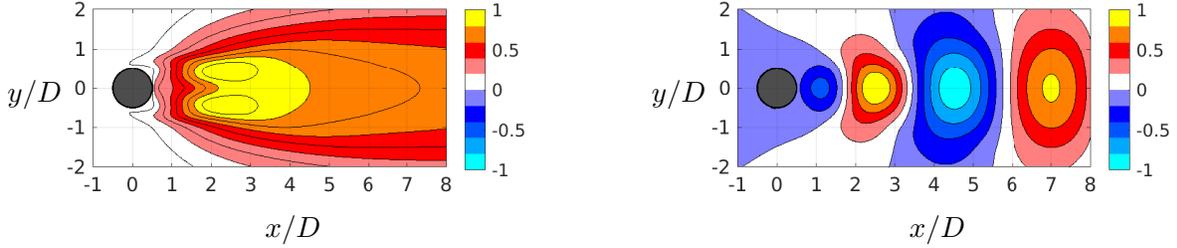
\begin{figure}[h!]
\centering

\subfloat[Norm of the absolute value of the first DMD mode, $N_1$ ]{
	\begin{tikzpicture}
		\node (img) { \includegraphics[width=0.45\textwidth] {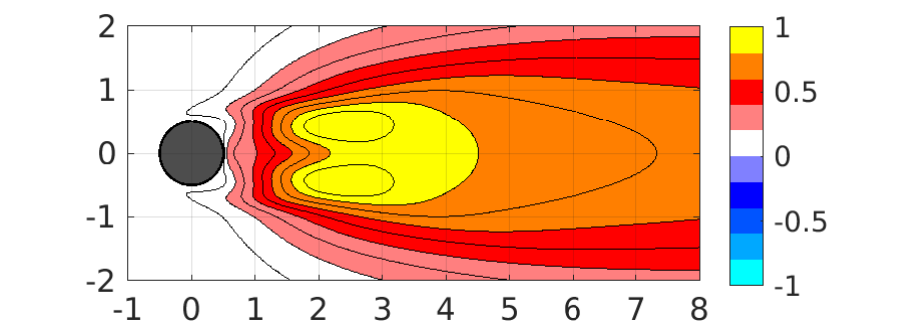} };
	 	\node[below=of img, node distance=0cm, yshift=1cm] {$x/D$};
	 	\node[left=of img, node distance=0cm, rotate=0, anchor=center, yshift=0cm, xshift=1.4cm] {$y/D$};
		 \label{subfig:Cyl_N1}
	\end{tikzpicture}}
	\hfill
\subfloat[Real part of the normal component of the velocity of the first DMD mode, $\mathcal{R}(u_y^1)$]{
	\begin{tikzpicture}
		\node (img) { \includegraphics[width=0.45\textwidth] {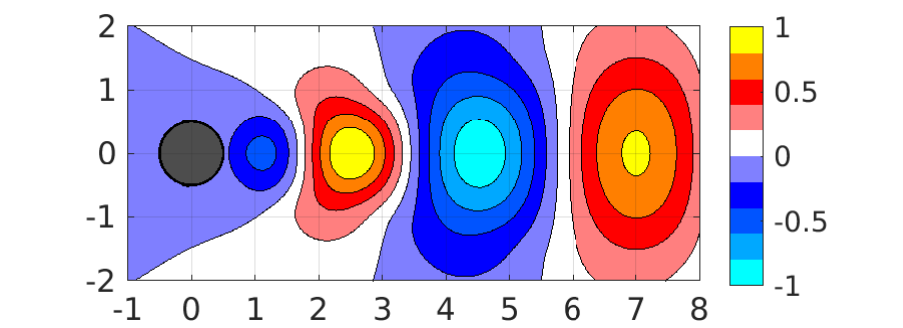}  };
		\node[below=of img, node distance=0cm, yshift=1cm] {$x/D$};
	 	\node[left=of img, node distance=0cm, rotate=0, anchor=center, yshift=0cm, xshift=1.4cm] {$y/D$};
		\label{subfig:Cyl_Vy1}
	\end{tikzpicture}} 
	\caption{a) $N_1$ and b) $\mathcal{R}(u_y^1)$ of the first DMD mode, corresponding to frequency $f_1$.}
\end{figure}

As the number of clusters or retained principal components increases, the characteristic lobe shapes become less visible, with the domain being divided into more granular regions, including those away from the central wake. Moreover, the structure of this optimal clustering solution shows a strong resemblance to the spatial distribution of the first Dynamic Mode Decomposition (DMD) mode, specifically the norm of its magnitude, defined as $N_1 = \sqrt{ {|u_x^1|}^2 + {|u_y^1|}^2 }$ (Fig.~\ref{subfig:Cyl_N1}). This DMD mode is known to represent the global wake motion and is closely related to the fundamental flow instability, further reinforcing the relevance of the clustering results.


\begin{figure}[h!]
\centering
	\includegraphics[width=0.45\textwidth] {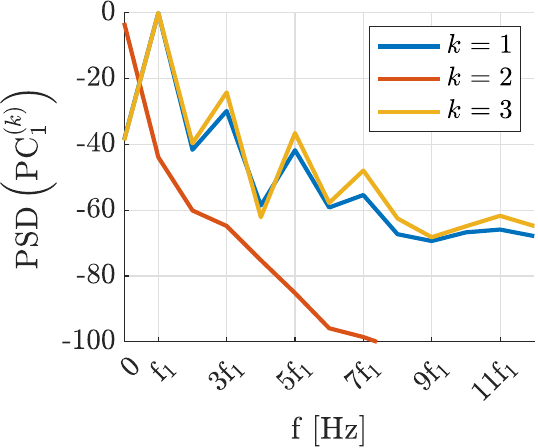}  
	\caption{Power Spectral Density of the first principal component (PC$_1$) for each cluster $k$, $\text{PSD} \left(\text{PC}_1^{(k)}\right)$, of the optimal solution.}
	\label{fig:PSD}
\end{figure}

Figure~\ref{fig:VQPCA_k_v}, which illustrates the clustering solution for $u_y$, supports the observation that the cluster structures change significantly as the number of clusters increases. This analysis focuses on the optimal case with $k=3$, where the domain is divided into wake regions that closely resemble the configuration shown in Fig.\ref{fig:2D60y100}(d). The similarity becomes even more evident when the clustering is compared with the real part of the first DMD mode of $u_y$ (Fig.\ref{subfig:Cyl_Vy1}), which is associated with the oscillation frequency of the wake. In particular, the clustering solution is symmetric, reflecting the inability of the algorithm to distinguish between regions above and below the cylinder along the normal direction ($y$).


To further validate the correlation between the clustering solution and the wake, we compute the power spectral density (PSD) of the principal component for each cluster $k$, denoted $\text{PSD} \left(\text{PC}_1^{(k)}\right)$. The results are shown in Figure~\ref{fig:PSD}. The frequency of the first DMD mode, denoted as $f_1$, corresponds to the wake oscillation frequency. The spectral density of clusters $1$ and $2$ peaks at $f_1$, indicating a strong correlation with this mode. In contrast, the principal component of cluster $3$ shows a maximum spectral density at $f=0$, revealing a steady region.

These results demonstrate that the Davies-Bouldin Index (DBI) provides an optimal solution closely related to the flow physics. It effectively identifies (i) regions that correspond to the structural sensitivity of the first instability in the wake, particularly with respect to the streamwise velocity component, and (ii) the first DMD mode, which is associated with the wake behind the cylinder generated by this instability. For consistency in the analysis and subsequent comparisons, $k=3$ is fixed from this point onward.



\subsubsection{Robustness assessment of the method and its calibration}\label{subsec:cyl_comp}

Having selected all parameters based on a reference case, the next step is to assess the robustness and reliability of both the method and its calibration. To this end, clustering results are compared in the three cylinder datasets (\textit{2D60}, \textit{2D100}, and \textit{3D280}), and the main assumptions underlying the method are evaluated. Specifically, the clustering procedure is shown to extract physically meaningful features from the flow. In particular, clustering based on the streamwise velocity component, $u_x$, was found to reveal regions potentially associated with structural sensitivity zones.


\begin{figure}[h]
  \centering
  \subfloat[Streamwise component of the velocity, $u_x$] {\includegraphics[height=0.31\textheight]{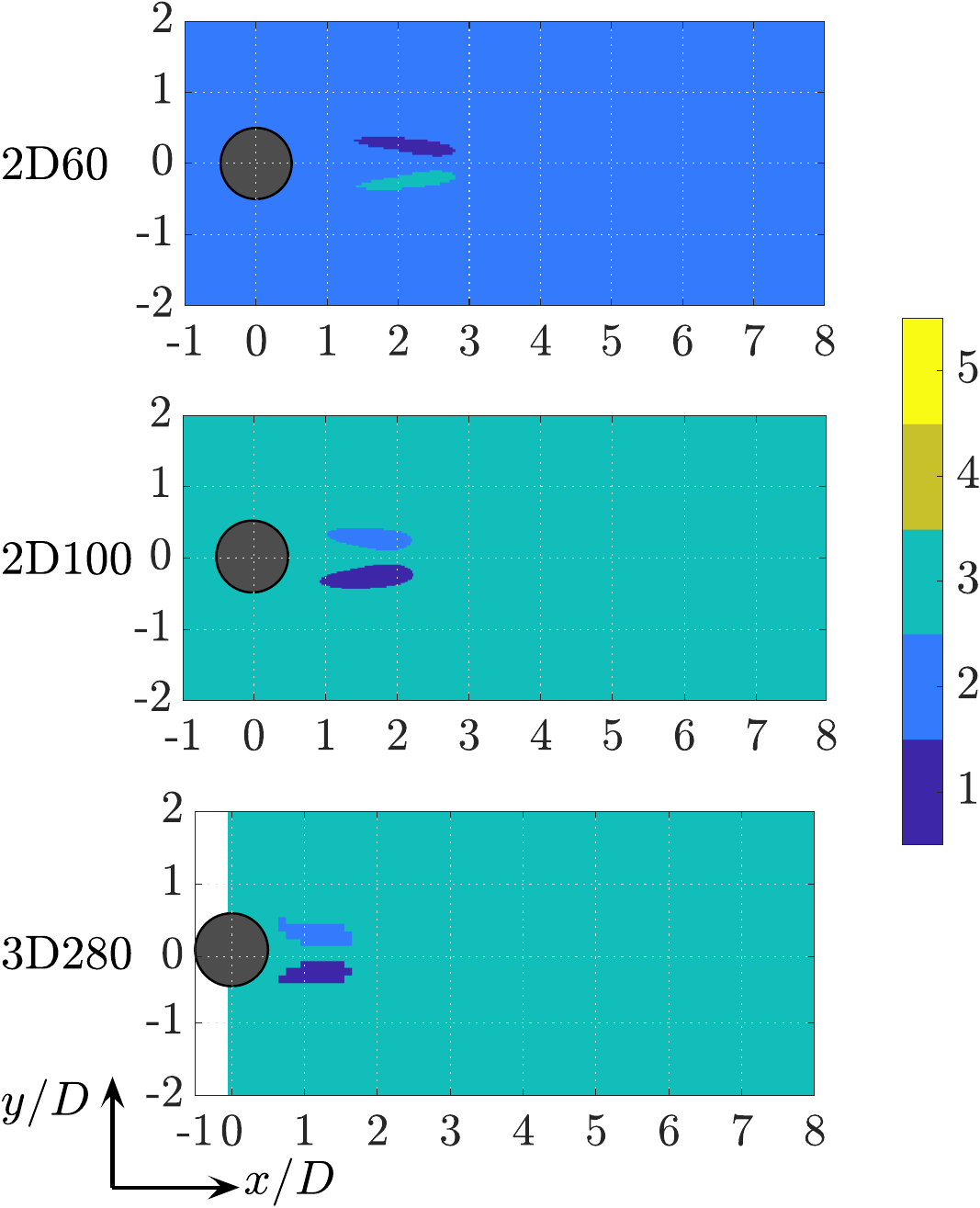}} \hspace{1cm}
  \subfloat[Normal component of the velocity, $u_y$]
  {\includegraphics[height=0.3\textheight]{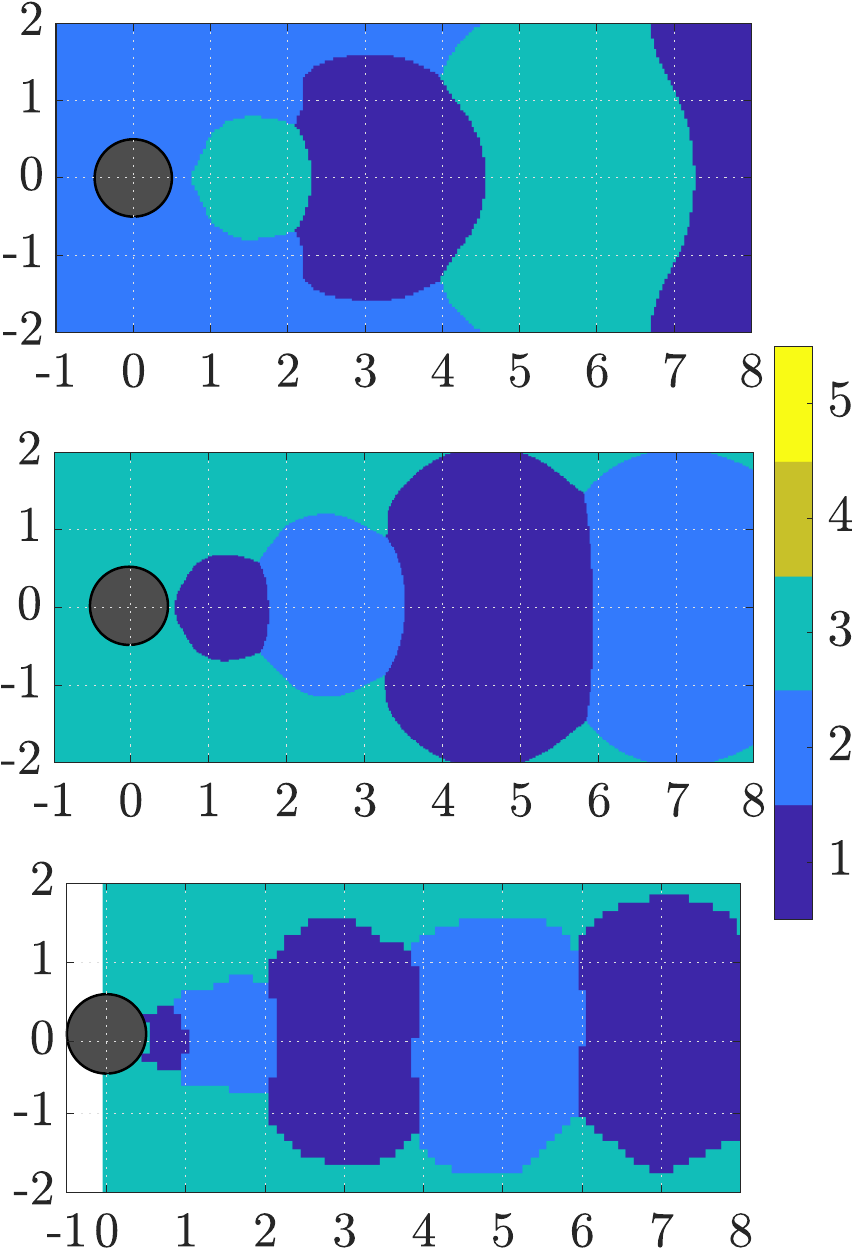}}
  \caption{Clustering solution on a) $u_x$ and b) $u_y$, in time in the three cylinder datasets.}
  \label{fig:VQPCA_comp_cyl}
\end{figure}

The comparison described above is illustrated in Fig.\ref{fig:VQPCA_comp_cyl}, which shows the clustering results for the streamwise and normal velocity components across the three cases. The same hyperparameter values used in the previous analysis are maintained, namely $k=3$ and $n_t$ as listed in Table~\ref{tab:nt_selected}. For the three-dimensional case, results are shown on the mid-plane in the $x$-$y$ direction. In all cases, the characteristic lobes previously identified in \textit{2D100} are present, and their structure aligns well with existing literature~\cite{Giannetti2007, Strykowski1990}. Moreover, as the Reynolds number increases, the lobes become smaller and shift closer to the cylinder, which is consistent with experimental observations (see Fig.20 in Ref.\cite{Strykowski1990}). These findings suggest that clustering via VQPCA successfully identifies regions of heightened structural sensitivity in the nonlinear flow solution and captures their evolution with Reynolds number.


The lower spatial resolution of the \textit{3D280} dataset results in less defined cluster contours in Fig.\ref{fig:VQPCA_comp_cyl}. However, the algorithm yields consistent results across all cases, despite variations in Reynolds number, mesh characteristics (including domain extent, discretization, and spatial dimensionality), and the inclusion of the third velocity component. Additional analysis of the three-dimensional case is provided in Appendix~\ref{sec:3Dclust}, where it is shown that clustering primarily captures a two-dimensional wave structure, confirming that the optimal clustering solution remains effectively bi-dimensional.


\section{Application on the flow of two synthetic jets}\label{sec:Results_Jets}

Having established the robustness of the clustering framework on a canonical problem, we now turn to a more complex, unsteady configuration to assess its applicability beyond globally periodic wakes.

\subsection{Flow and instabilities on two synthetic jets}\label{sec:Jets_instabilities}
Synthetic jets are devices of significant industrial relevance and their operation has been extensively investigated in the literature~\citep{Kraletal1997,LeClainchePerez2020}. A synthetic jet typically consists of a cavity with a membrane or piston at one end and an orifice at the other. The sinusoidal motion of the piston expels and entrains fluid through the orifice, forming a train of counter-rotating vortices that generate thrust.

The flow induced by two synchronized synthetic jets represents a particularly interesting configuration, as the interaction between the two jets can trigger a global instability that breaks the flow symmetry~\citep{MunozLeClainche2022}. The resulting flow dynamics are governed by two non-dimensional parameters: the Reynolds number, defined as $\mathrm{Re} = \frac{U D}{\nu}$, where $U$ is a characteristic velocity, $D$ is the diameter of a jet orifice, and $\nu$ is the kinematic viscosity of the fluid; and the Strouhal number, defined as $St=\frac{fD}{U}$, where $f$ is frequency of the piston oscillation.

The frequency and time $t$ define the periodic phase $\varphi = 2 \pi f t \in [0,2\pi]$, which distinguishes the injection ($\varphi \in [0,\pi]$) and suction ($\varphi \in [\pi, 2\pi]$) parts of the cycle. According to Ref.~\cite{MunozLeClainche2022}, a symmetry-breaking instability emerges at Reynolds numbers in the range of $140 - 150$ for $St = 0.03$ in the two-dimensional configuration.




\begin{figure}[h]
  \centering
  \subfloat[Injection, $\varphi=3\pi/4$] {\includegraphics[width=0.43\textwidth]{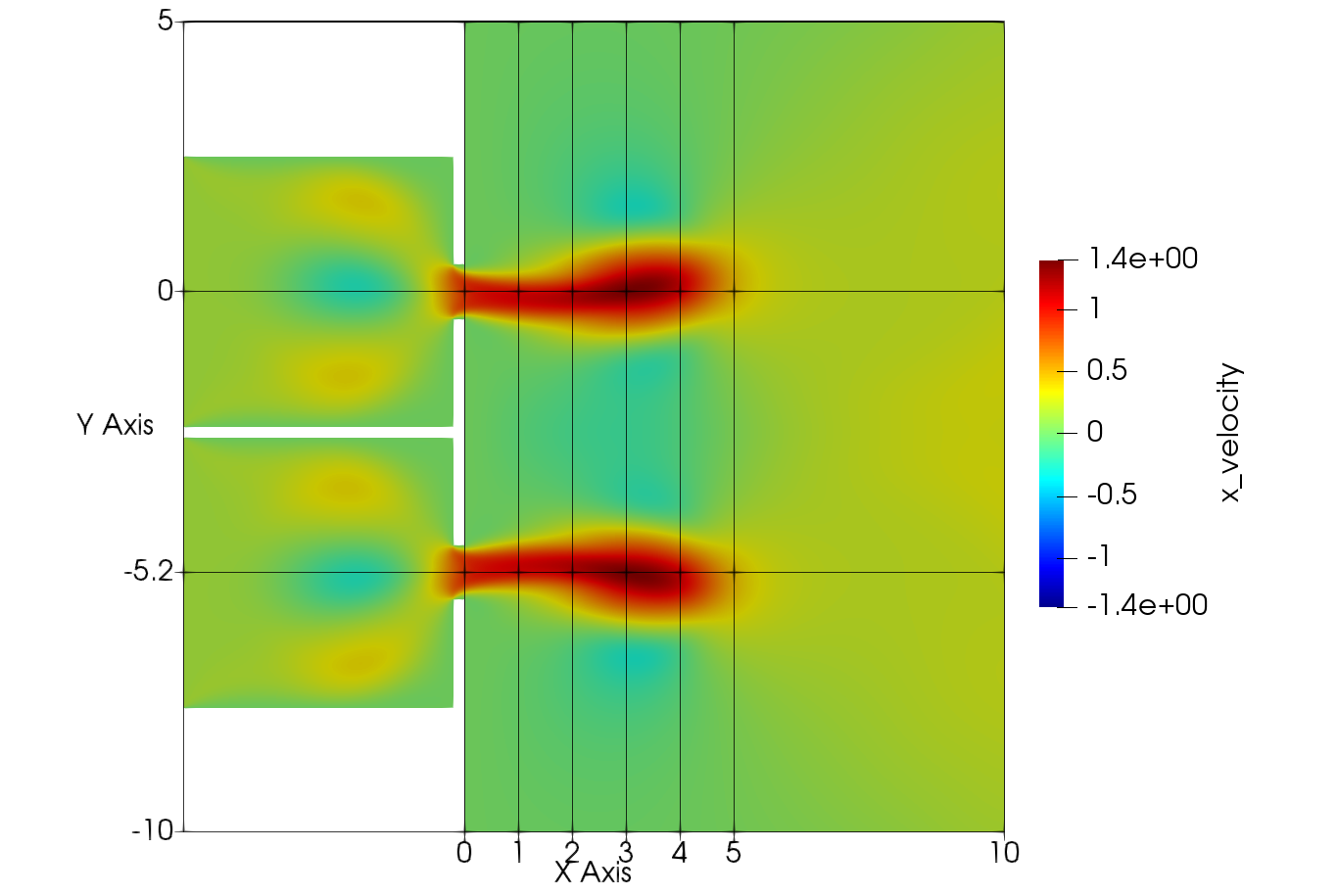}\label{subfig:Re100_inj}} \hspace{0.5cm}
  \subfloat[Suction, $\varphi=3\pi/2$] {\includegraphics[width=0.43\textwidth]   {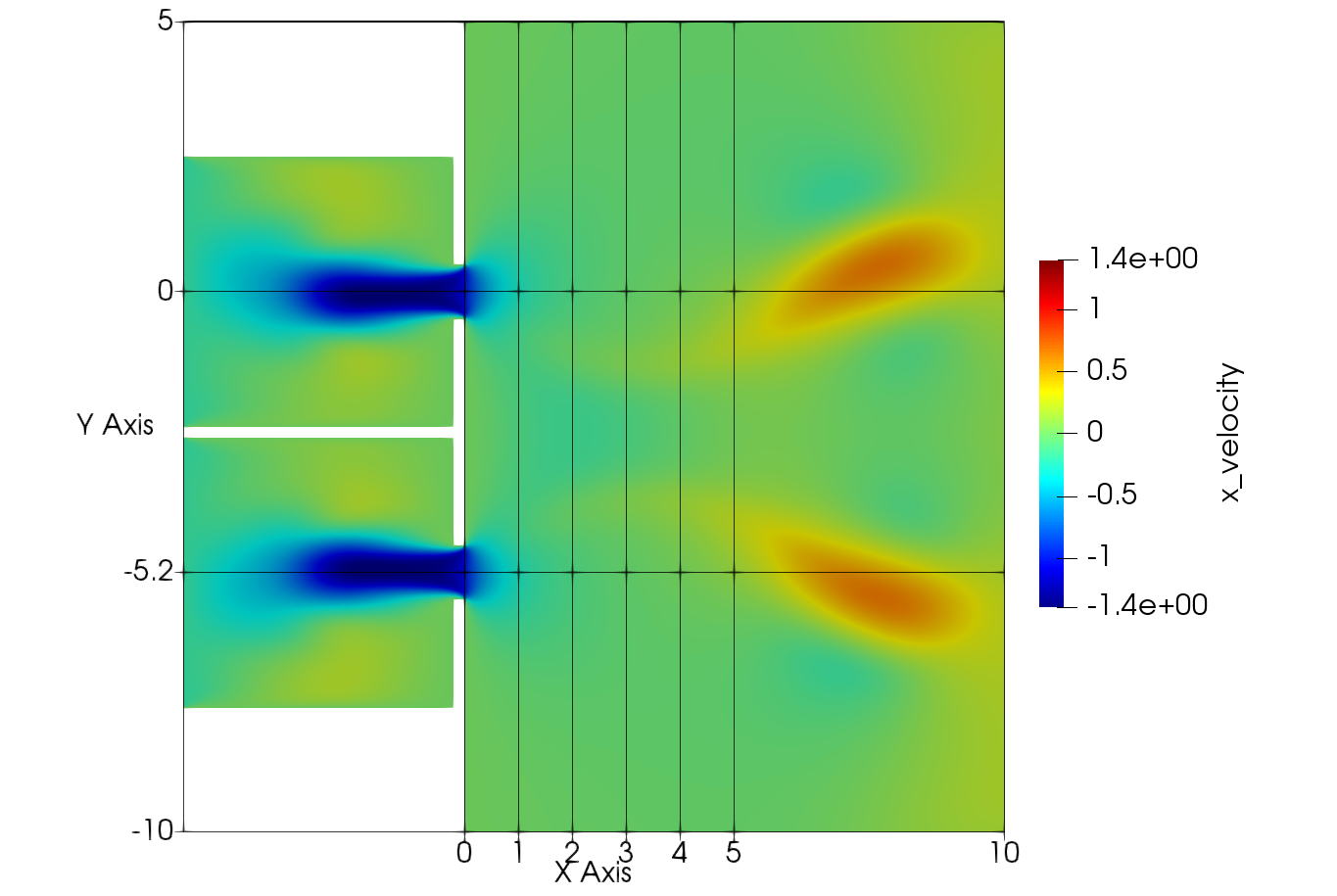}\label{subfig:Re100_suct}}
  \caption{Contours of the streamwise component of the velocity, $u_x$, of the case at $\mathrm{Re}$=100 in a characteristic time during the a) injection and b) suction phases.}
  \label{fig:Re100_vx}
\end{figure}

Figure~\ref{fig:Re100_vx} presents velocity contours in two representative phases of the actuation cycle for a symmetric flow at $\mathrm{Re} = 100$, which illustrates the jet mechanism. During the injection phase (Fig.\ref{subfig:Re100_inj}), fluid is expelled from the cavities through the orifices, generating distinct flow streams that propagate downstream. For simplicity, these are referred to as jet streams. In contrast, during the suction phase (Fig.\ref{subfig:Re100_suct}), the flow is drawn back into the cavities, while the vortices formed during the injection phase continue their downstream evolution.

\begin{figure}[htb]
  \centering
  \includegraphics[width=0.9\textwidth]{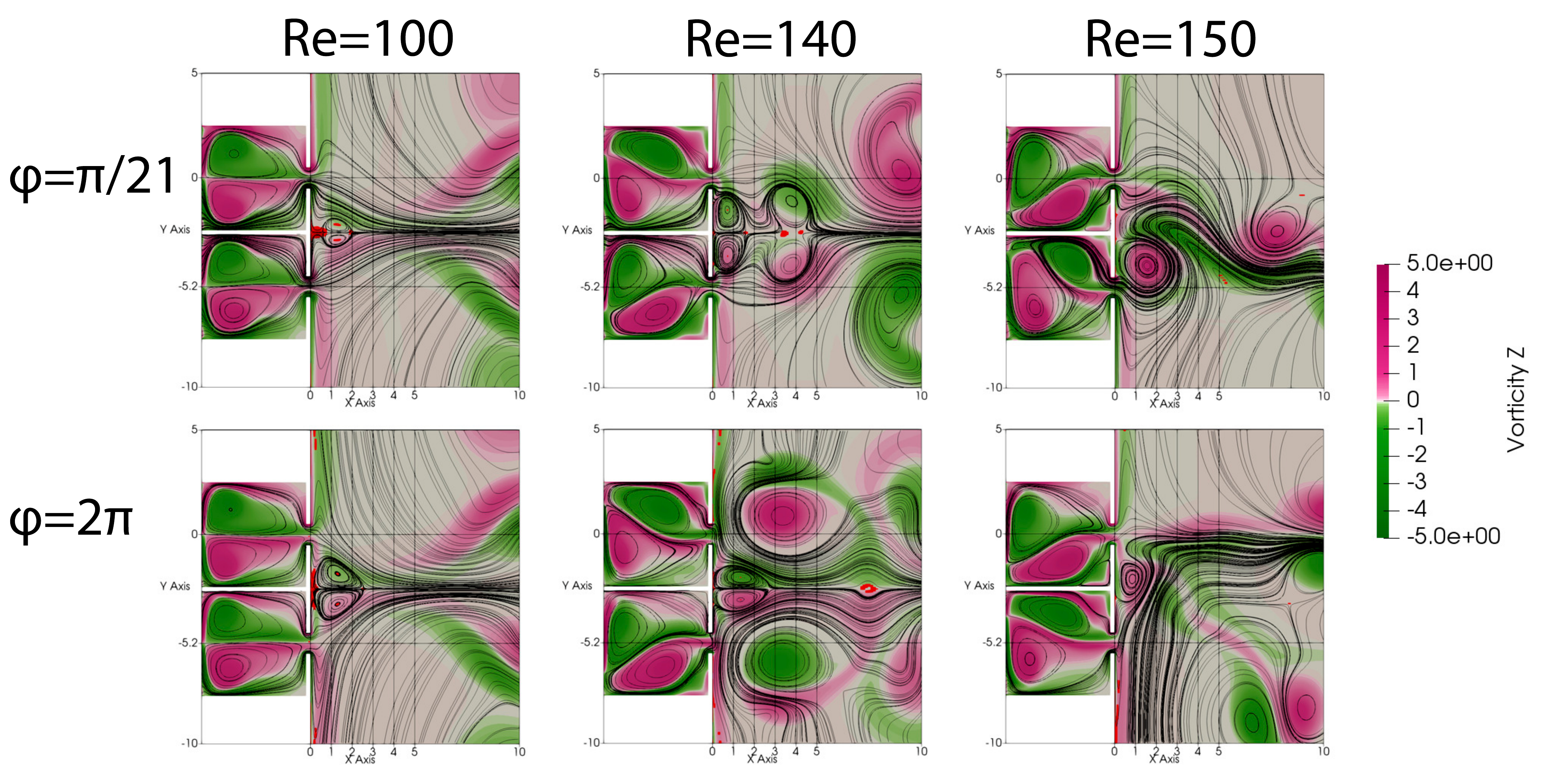}
  \caption{For each sub-figure and each Reynolds number, streamlines and vorticity contours of snapshots $\varphi=\pi/21$ and $\varphi=2\pi$, corresponding to the beginning of the injection phase and the end of the suction phase respectively. The red dots correspond to saddle points.}
  \label{fig:Jets_vort}
\end{figure}

The effects of the flow instability are illustrated in Figure~\ref{fig:Jets_vort}, which displays vorticity contours and streamlines at two distinct time instants for Reynolds numbers $\mathrm{Re} = 100$, $140$, and $150$. The flow remains symmetric at the lower Reynolds numbers but becomes clearly asymmetric at $\mathrm{Re} = 150$, indicating the onset of instability. Under symmetric conditions, two recirculation bubbles form between the jets and the symmetry line—features that play a crucial role in the development of the instability, as discussed in the following sections. These bubbles reach their maximum lengths at $\varphi = 2\pi$, measuring approximately $2.5D$ and $3D$ in the $x$-direction for $\mathrm{Re} = 100$ and $140$, respectively. Additionally, their temporal persistence differs: at $\mathrm{Re} = 140$, the bubbles are present during the phase range $\varphi \in [37\pi/42, 5\pi/21]$, while at $\mathrm{Re} = 100$ they exist for a shorter interval, $\varphi \in [41\pi/42, 2\pi/21]$.


The symmetric and asymmetric flow regimes significantly influence the downstream field, potentially limiting the performance and applicability of synthetic jet devices. Therefore, controlling the onset of the symmetry-breaking instability becomes essential. To this end, the clustering method developed in Sec.~\ref{sec:Results_Cyl} is applied to the present configuration with the objective of identifying the potential structural sensitivity zones responsible for triggering the instability.


\subsection{Dataset}\label{sec:Jets_dataset}
The dataset analyzed corresponds to two synchronized synthetic jets operating at a Strouhal number $S_t=0.03$ and Reynolds numbers $Re=100$, $Re=140$ and $Re=150$. The flow is governed by the two-dimensional, incompressible Navier–Stokes equations (Eqs.\ref{eq:cont} and \ref{eq:NS}), non-dimensionalized with the characteristic velocity $U=1$ and length $L=D=1$. At the inlet, a sinusoidal axial velocity profile mimics the motion of the jet piston, prescribed as $u_x = \frac{1}{5} \sin (2 \pi f t) $ and $u_y = 0$ along with Neumann boundary conditions for pressure. No slip boundary conditions ($u_x=u_y=0$) are imposed on all walls, while the outflow condition follows the formulation proposed by Ref.\cite{Dong2014}.


The computational domain spans $x\in [-5.2D, 300D]$ in the streamwise direction and $y \in [-245.7D , 240.5D]$ in the normal direction, with the origin of the coordinate system located at the upper jet nozzle. The mesh consists of $3,182$ macro-elements, each one discretized using a spectral element method with a polynomial order of $18$ in both directions, yielding a total of $1,030,968$ grid points. Original flow simulations were performed using the open source spectral element solver \texttt{Nek5000}\citep{nek5000}, as reported in\cite{MunozLeClainche2022}, where further computational details can be found. In addition to the original datasets, new simulations were performed in the present study to explore flow control strategies. These modifications and their corresponding configurations are discussed in Sec.~\ref{sec:Jets_FlowControl}.


Clustering is performed on the dataset corresponding to $\mathrm{Re}=100$, and subsequently the results are extrapolated to the cases with higher Reynolds numbers. The dataset includes $N_v = 2$ flow variables, namely, the two velocity components, sampled at $N_x = 45$ and $N_y = 22$ equispaced grid points in the streamwise and normal directions, respectively. This corresponds to a spatial domain of $x \in [0.9D, 35.6D]$ and $y \in [-10.3D, 5.7D]$. A total of $N_t = 4369$ snapshots were collected over $7$ saturated piston cycles (from the $24^{th}$ to the $30^{th}$ cycle), with $624$ snapshots per cycle, providing a high-resolution temporal representation of the periodic dynamics.


\subsection{Data preparation}\label{sec:Jets_dataprep}
Based on the clustering results obtained for the cylinder flow in Sec.\ref{sec:Cyl_clus}, the data matrix $\bX \in \mathbb{R}^{n \times p}$ is constructed as schematized in Fig.\ref{fig:scheme_X}, where $n = n_v N_{xy}$ and $p = n_t$. We select $n_v = 2$ flow variables, namely, the velocity components $u_x$ and $u_y$, and $n_t = 42$ snapshots, corresponding to one complete cycle. These snapshots are uniformly sampled in time over the 24$^{\mathrm{th}}$ cycle. This selection balances temporal resolution with computational efficiency, as increasing $n_t$ significantly increases the cost of the clustering process.

\subsection{Clustering results}\label{sec:Jets_clust}
The application of this technique to the flow generated by two synthetic jets at $\mathrm{Re}=100$ was originally introduced by Muñoz et al. in Ref.~\cite{MunozDaveetal2023} (Fig. 18). Key flow features such as jet streams, their lateral separation and their eventual merging into a single jet in downstream evolution were successfully identified. This symmetric configuration is selected as the reference case.

\begin{figure}[htb]
  \centering
    \subfloat[DBI for \(k\in {[2, 8]}\)  ]{\includegraphics[height=0.15\textheight]{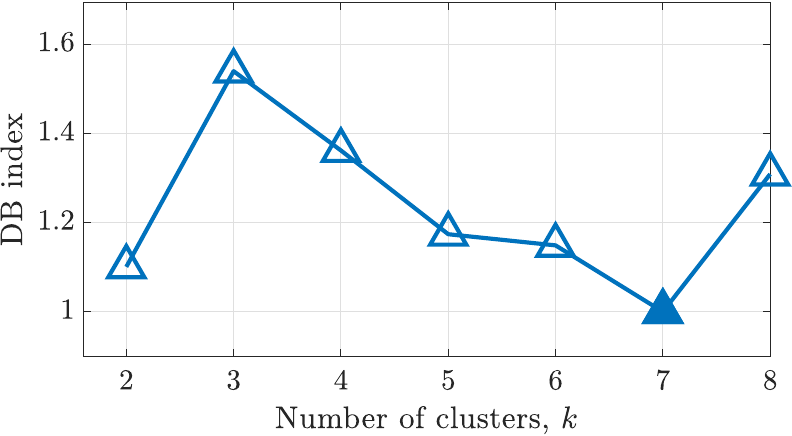}\label{subfig:DBI_jets}} \hspace{1cm}
  \subfloat[Clustering contours on the optimal solution, $k$=7]  {\includegraphics[height=0.15\textheight]{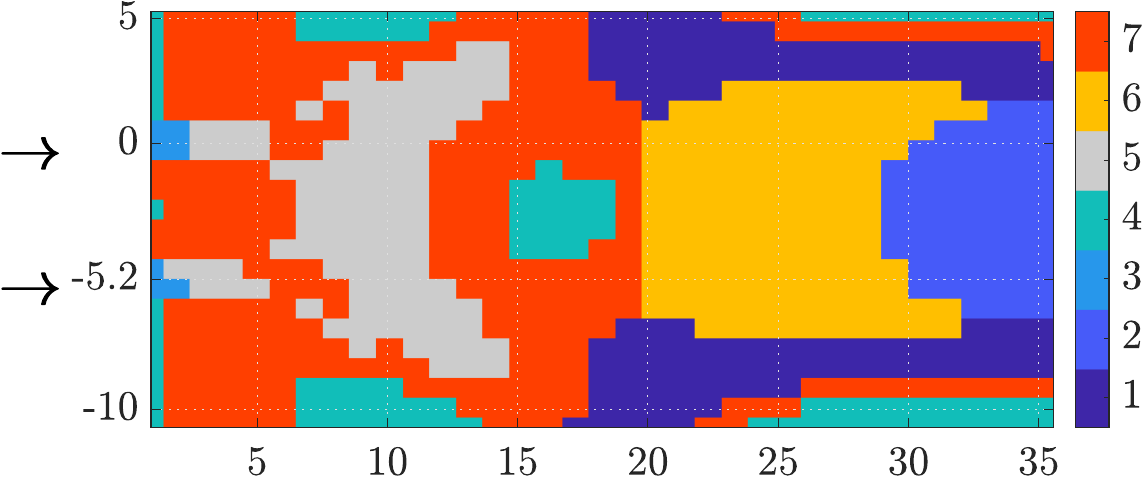}\label{subfig:VQPCA_jets}}
  \caption{a) Davies Bouldin index for the clustering results with $k\in {[2, 8]}$. The optimal solution based on the DBI is highlighted with a filled marker. b) Contours of the clusters identified for the optimal clustering, k=7. Both figures correspond to synthetic jets flow at  $\mathrm{Re}$=100.}
  \label{fig:DBI_clust_jets}
\end{figure}

In the present study, clustering is carried out following the methodology described in Section~\ref{subsec:hyperp} to identify structural sensitivity zones. In particular, clustering solutions for a range of cluster numbers, $k \in [2,8]$, are evaluated. Figure~\ref{subfig:DBI_jets} presents the corresponding Davies–Bouldin index (DBI) values, with the optimal solution at $k = 7$ (indicated by the lowest DBI) highlighted using a filled marker. The resulting cluster contours for the streamwise velocity component in the optimal case are shown in Fig.~\ref{subfig:VQPCA_jets}.

\begin{figure}[htb]
  \centering
  \includegraphics[width=\textwidth]{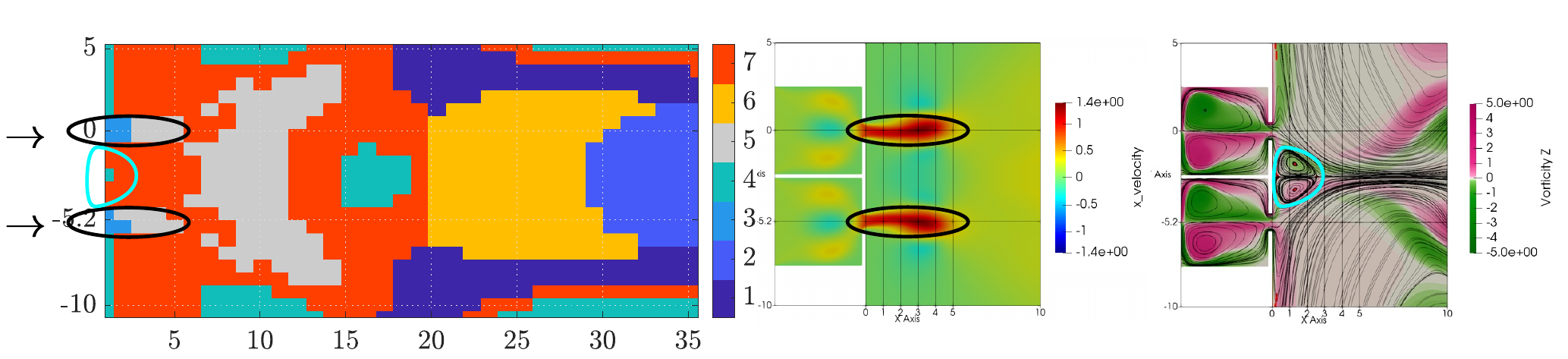}
  \caption{Identification of the jet streams and recirculation bubbles, black and blue lines respectively, from the clustering solution of Fig.~\ref{subfig:VQPCA_jets}.}
  \label{fig:VQPCA_identif}
\end{figure}

A more detailed analysis of the cluster solution is now conducted, focusing specifically on the region near the jet exits. Figure~\ref{fig:VQPCA_identif} highlights this area and marks the locations of two key flow features: jet streams and recirculation bubbles, indicated by black and blue lines, respectively. Jet streams are accurately captured by cluster $3$ in the immediate vicinity of the orifices, and by cluster $5$ further downstream. Recirculation bubbles, which form between jet streams, are primarily associated with cluster $7$. Identification of these coherent structures, together with the demonstrated ability of the clustering algorithm to detect structurally sensitive regions in the wave-cylinder case (Sec.~\ref{sec:Cyl_clus}), suggests that both jet streams and recirculation zones may play a central role in the onset of flow instabilities.

To explore the practical implications of the identified sensitivity regions, additional simulations have been conducted with slight modifications (boundary conditions, computational domain, and mesh) to the baseline setup. These cases aim to illustrate how clustering results can guide flow control strategies, setting the basis for a more detailed study. In particular, localized perturbations in the form of point forces and passive disruptors are applied to selected clusters in order to assess their influence on the onset of symmetry breaking. The next section briefly summarizes illustrative results obtained when these regions are perturbed, while full details are deferred to Appendix~\ref{app:Jets_FlowControl}.




\subsection{Prospective flow control as clustering application}\label{sec:Jets_FlowControl}

Although already mentioned, the primary objective of this work is the identification of structurally sensitive regions using clustering, a secondary analysis explores the potential of these results to guide flow control strategies. The present section discusses the main trends found; detailed configurations and parametric variations are reported in Appendix~\ref{app:Jets_FlowControl}.



As a prospective application of the clustering framework, targeted perturbations were applied to selected clusters to assess their influence on the onset of symmetry breaking in the flow of two synthetic jets. The geometry and numerical setup follow the configurations presented in Ref.\cite{MunozLeClainche2022}, as summarized in Section\ref{sec:Jets_dataset}, with suitable modifications to accommodate the control mechanisms. Two strategies are then explored: point forces and disruptors. Point forces consist of localized, time-dependent body forces applied at specific locations in the flow, whereas disruptors are passive square elements introduced into the domain to modify the local flow topology through geometric obstruction. Both approaches are designed to investigate the sensitivity of the regions identified by clustering, rather than to achieve optimal control solutions.

The efficacy of the control strategies is assessed using a newly introduced indicator: the symmetry breaking cycle (cycle$_b$), which denotes the piston cycle at which the flow asymmetry becomes first clearly observable. For the destabilization objective at $\mathrm{Re}=140$, success is defined as achieving asymmetry within the first analyzed cycles. In contrast, the stabilization objective at $\mathrm{Re}=150$ aims to delay the onset of asymmetry beyond the original transition point, cycle${_b}_{org} = 22$. An effective strategy in this case would yield cycle$_b$ > cycle${_b}_{org}$, or ideally maintain a symmetric flow throughout.


Point-force perturbations were primarily used to test the sensitivity of jet-stream clusters. The main finding is that forces applied within clusters associated with the jet streams have a strong influence on the onset of symmetry breaking. In flows close to instability, such perturbations were able to either advance or delay the instability depending on their placement and orientation, confirming the high structural sensitivity of these regions. In particular, perturbations aligned with the dominant jet dynamics were found to accelerate symmetry breaking, while forces applied in neighboring clusters produced a moderate stabilizing effect. These trends, described in Appendix~\ref{app:Jets_FlowControl} (see Fig.~\ref{fig:Forces_all} for an overview), demonstrate that the jet-stream clusters identified by VQPCA correspond to dynamically active regions whose perturbation rapidly affects the global flow state.

Disruptors were introduced to investigate the sensitivity of the cluster associated with the recirculation bubbles between the two jets, which were found to contribute to flow stabilization. In contrast to point forces, disruptors consistently produced a strong stabilizing effect when placed within this cluster. In several cases, the onset of symmetry breaking was significantly delayed, and in the most favorable configurations the flow remained symmetric for several tens of cycles. Flow visualizations indicate that disruptors modify the shape and persistence of the recirculation bubbles, promoting a more symmetric configuration. A representative comparison of disruptor locations and their stabilizing effect is provided in Appendix~\ref{app:Jets_FlowControl} (see Fig.~\ref{fig:Cont_E}). These results highlight the central role of the recirculation-bubble cluster in governing the symmetry-breaking instability.

\section{Concluding remarks} \label{sec:Conc}

This work introduces an efficient, low-cost approach for identifying structural sensitivity in fluid flows, with computational times typically on the order of seconds. This efficiency is achieved through a data-driven method that relies exclusively on information from the direct problem, avoiding the computational expense of adjoint-based techniques.


The core of the method lies in a clustering algorithm based on Vector Quantization Principal Component Analysis (VQPCA), which partitions the flow domain using a low-dimensional representation of the system. Specifically, the method builds this space from a single temporal correlation, retaining only the dominant principal component for each cluster. This enables each cluster to be uniquely defined by its temporal behavior, effectively partitioning the flow into dynamically coherent regions. Significantly, the results suggest that small perturbations in these identified regions can trigger substantial changes in global flow dynamics, indicating a strong link between clustering-based partitioning and structural sensitivity.



The approach is validated through the canonical case of flow past a circular cylinder, a well-established benchmark. The results are consistent with known physical structures and the method was proved robust throughout variations in sampling resolution, Reynolds number, and dimensionality of the data set, demonstrating consistency with established results in the literature.


The methodology is further tested on a more complex scenario: the interaction of two planar synthetic jets. In this case, the clustering identified dynamically meaningful regions corresponding to the jet streams and the recirculation bubbles, both of which emerged as zones of potential structural sensitivity. These insights guided a prospective secondary application: the evaluation of flow control strategies based on perturbations targeted at these clusters. Although not the main focus of this study, these tests illustrate the ability of clustering to guide the placement of perturbations to either advance or delay symmetry breaking.

In summary, the clustering-based framework presented here is both robust and computationally efficient, providing rapid insight into structural sensitivity zones of fluid dynamic problems. Beyond the cases studied, the methodology shows strong potential for more complex scenarios, including the analysis of turbulent flows, experimental datasets, and integration with machine learning frameworks for real-time flow analysis and control.


\begin{acknowledgments}
EM, JAM, and SLC would like to thank the Comunidad de Madrid for its support through the Call Research Grants for Young Investigators from the Universidad Politécnica de Madrid and also acknowledge the grant PID2020-114173RB-I00 funded by the Spanish Ministry of Culture and Innovation MCIN / AEI/10.13039/501100011033. \\
HD and AP would like to thank the financial support received through the Individual Fellowship Call of Université Libre de Bruxelles (ULB) and the European Commission Grant Agreement number 801505 within the framework of the Marie Sklodowska-Curie Actions (H2020) program.
The research was sponsored by the Fédération Wallonie-Bruxelles, via "Actions de recherche concertées - projets avancés" for 2022-2027.
\end{acknowledgments}

\bibliography{bibliography}
\appendix
\FloatBarrier
\section*{Appendices}
\section{Three-dimensional analysis of the clustering results on the cylinder} \label{sec:3Dclust}

This section presents a three-dimensional analysis of the clustering results for the \textit{3D280} case of flow past a circular cylinder. The goal is to examine how the clustering behaves along the spanwise direction and to assess whether these variations affect the interpretation of the two-dimensional results. To this end, the optimal clustering configuration ($k=3$, $q=1$) is compared against a higher-resolution case using more clusters and retained components ($k=4$, $q=4$).

\begin{figure}[hbt] 
  \centering
    \subfloat[Optimal hyperparameters, $k$=3, $q$=1] {\includegraphics[width=0.48\textwidth]{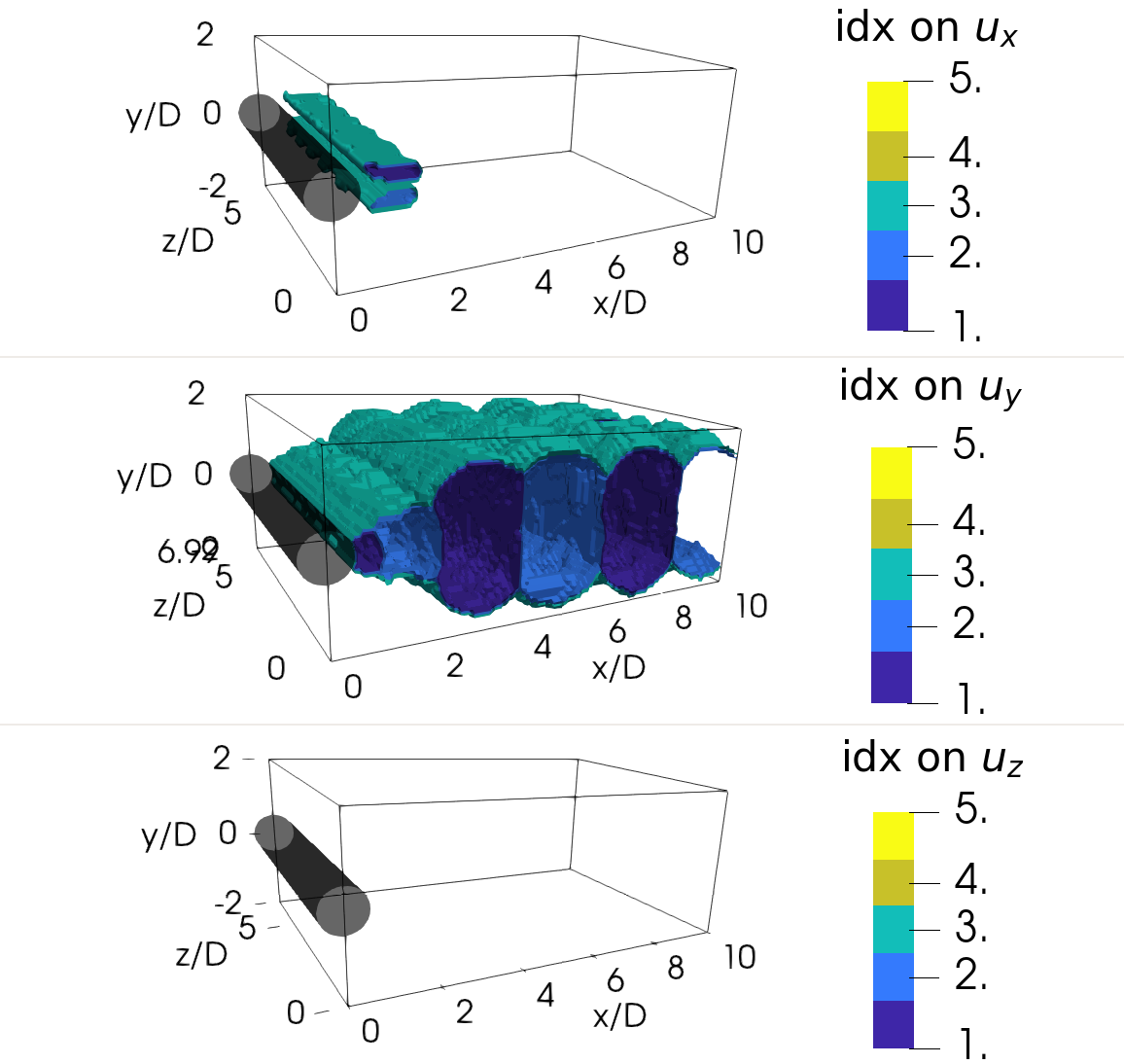}\label{subfig:VQPCA_3D_31}}   
    \subfloat[hyperparameters $k$=4, $q$=4] {\includegraphics[width=0.48\textwidth]{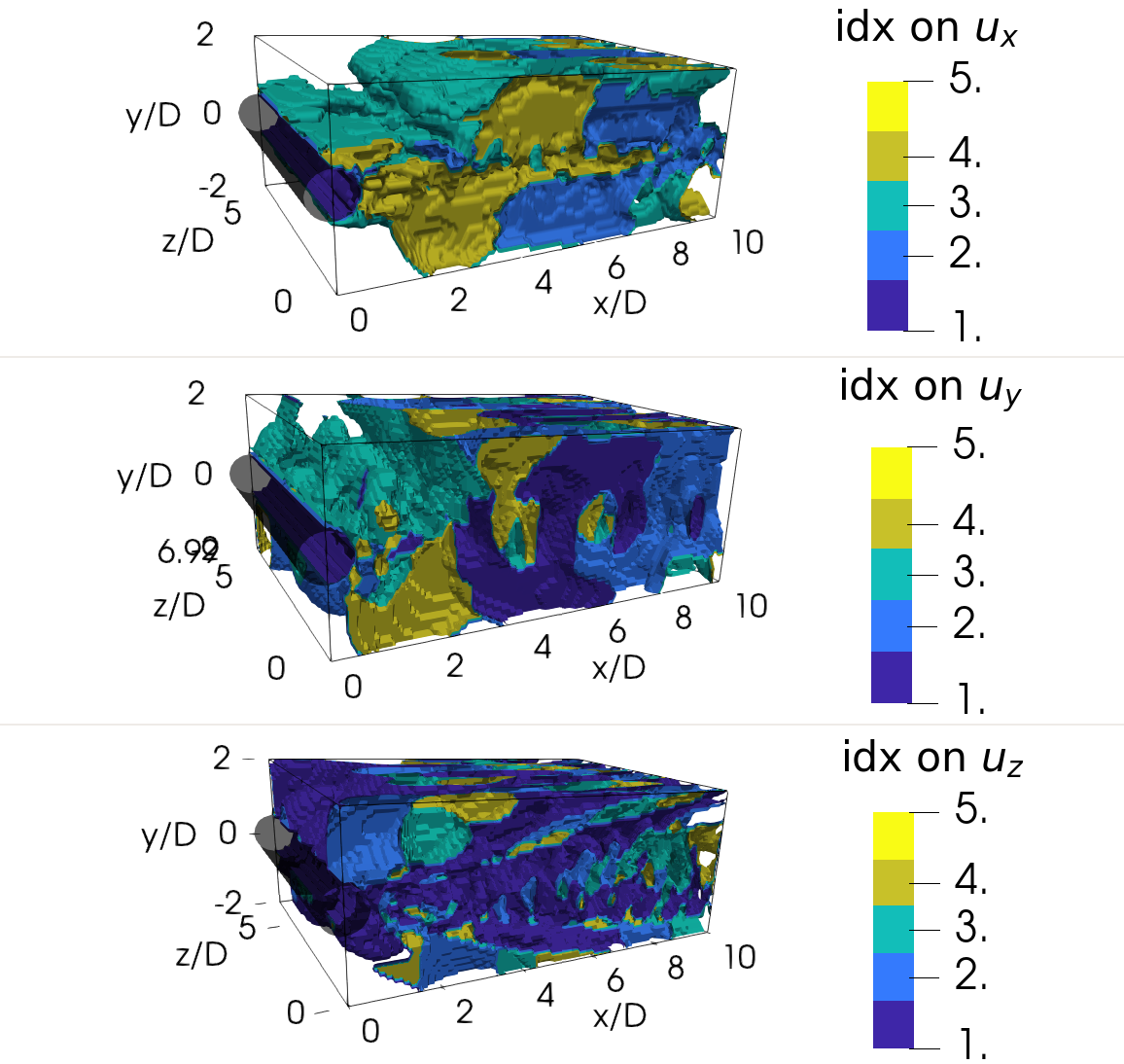}\label{subfig:VQPCA_3D_44}} 
    \caption{Isosurfaces of the clusters identified on \textit{3D280} corresponding to the variables $u_x$, $u_y$ and $u_z$ (in rows).}
  \label{fig:VQPCA_3D_both}
\end{figure}

Figure~\ref{subfig:VQPCA_3D_31} shows a three-dimensional view of the clusters for the optimal configuration. The clustering, applied to the streamwise and normal velocity components ($u_x$ and $u_y$), reveals the presence of distinct lobes and wavelike structures that persist throughout the spanwise extent. In particular, the spanwise component $u_z$ is not visualized, as the algorithm assigns all its grid points to a single cluster (cluster 3). This indicates that $u_z$ contributes minimally to the dominant temporal patterns captured by the model and that the clustering distribution is effectively two-dimensional.

Cluster 3 also dominates the far-field region in $u_x$ and $u_y$, suggesting that it corresponds to the most steady part of the domain. In contrast, clusters 1 and 2 are concentrated near the cylinder, where the primary flow dynamics is active. This behavior aligns with the results presented in Section~\ref{sec:Cyl_clus} and the reference spectral analysis (Fig.~\ref{fig:PSD}). Since the model retains only one principal component, it captures the largest-scale flow features. Consequently, clusters 1 and 2 encapsulate the dominant two-dimensional wave motion that drives the instability, while cluster 3 encompasses the background flow with lower temporal variability.


This interpretation is further supported by the clustering solution obtained for the same three-dimensional case using higher values for the hyperparameters. Figure~\ref{subfig:VQPCA_3D_44} shows the clustering results for $k=4$ and $q=4$, visualized as isosurfaces for all three velocity components. Compared to the optimal configuration, this higher-resolution setup allows the algorithm to capture finer spatial structures and provides a more detailed representation of the flow dynamics.

However, this increased resolution comes at the cost of interpretability. The isosurfaces become more complex and the previously clear associations with large-scale flow features are more difficult to discern. Notably, with $k=4$ and $q=4$, the clustering incorporates significant contributions from all three velocity components, leading to a fully three-dimensional cluster distribution.

These observations confirm that the clustering algorithm can effectively extract physically meaningful flow structures at different scales. When using low values of $k$ and $q$, the model captures the dominant large-scale features, such as the two-dimensional wave driving the instability. Increasing $k$ and $q$ allows the method to resolve smaller-scale structures and spanwise variations, offering flexibility depending on the resolution and analysis needs. Importantly, the optimal configuration identified in the main study ($k=3$, $q=1$), selected based on the Davies–Bouldin index, provides a clear and interpretable partitioning aligned with the primary instability mechanisms, while higher-resolution cases offer a path toward more detailed, three-dimensional flow analysis.

\section{Potential flow control as application of clustering techniques}\label{app:Jets_FlowControl}
This section provides a detailed description of the prospective flow control analysis summarized in the main text. The objective of this study is not to design or optimize control strategies, but to further illustrate the relevance of the structurally sensitive regions identified by the clustering approach when subjected to localized perturbations.

Two control mechanisms are examined in detail: point forces and disruptors. For each case, the corresponding configurations, their location relative to the clustering solution, and the resulting impact on the onset of symmetry breaking are presented in depth. This material complements the main text and suggests promising directions for clustering-guided flow control.

\subsubsection{Flow control with point forces}
Point forces are introduced as additional boundary conditions and applied symmetrically with respect to the centerline between the two jets. The spatial location of each force is defined by its streamwise position $x_f$, measured along the $x$-axis, and its normal position $y_f$, measured from the jet exits either towards the region between the jets (labeled as inside, I) or away from them (outside, O). The magnitude of the force, $|\boldsymbol{f}|$, is expressed as a percentage of the characteristic velocity $U$.

\begin{figure}[h!]
 	\centering
	\includegraphics[width=0.29\textwidth]{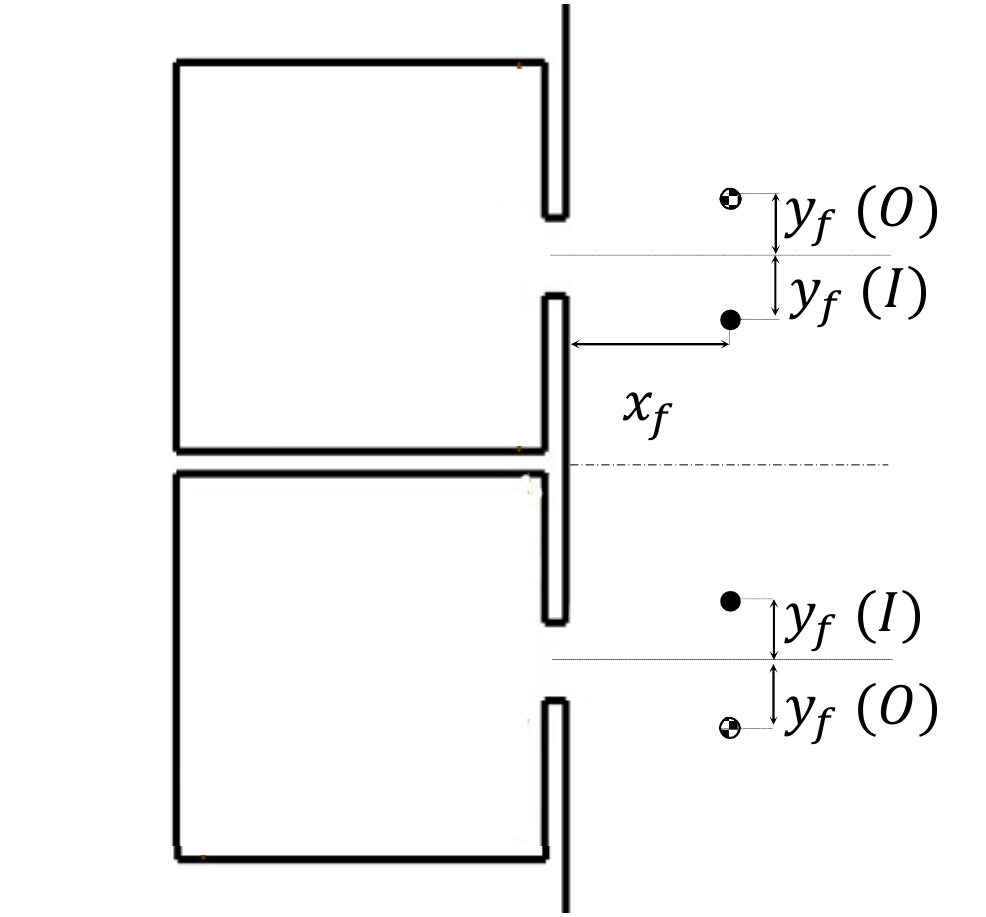} 
	\captionof{figure}{Scheme of the parameters that define the position of the forces.}
	\label{fig:Sch_F} 
\end{figure}

Forces are applied in different directions, including positive and negative streamwise orientations, denoted `$+x$' and `$-x$', respectively. Additionally, sinusoidal forcing—denoted as `$\sin$'—is considered, with the same frequency as the piston motion but in opposite phase, in an effort to more effectively counteract the natural oscillatory flow. This control input is defined as $\boldsymbol{f} = f_x \boldsymbol{i} = - |\boldsymbol{f}| \sin(2\pi f t), \boldsymbol{i} $. A schematic illustration of the positioning and force configuration is shown in Fig.~\ref{fig:Sch_F}.

Table~\ref{tab:Sim_A-F} summarizes the characteristics of the applied control forces. Cases A01 and A02 correspond to simulations at $\mathrm{Re}=140$, where forces are applied at $x_f = 3D$ and $y_f = 0D$ with the goal of destabilizing the flow. These forces have a magnitude of $|\boldsymbol{f}| = 10\%$ of $U$ and are oriented in the negative ($-x$) or positive ($+x$) streamwise direction.

In contrast, forces F01–F11 target stabilization of the flow at $\mathrm{Re}=150$. These are applied at $x_f = 1D$ and at various normal positions $y_f \in [0.25D \text{ (O)}, 1D \text{ (I)}]$. The first set, F01–F04, explores the effects of force direction ($+x$ or $-x$) and placement (inside or outside the jet region). Based on the most effective case, F03, further refinements are tested: (i) the force magnitude is varied in cases F05–F07, (ii) the vertical position $y_f$ is adjusted in F08 and F10, and (iii) sinusoidal forcing ($\sin$) is implemented in F09 and F11.

\begin{table}[h!]
  \centering
  \begin{tabular}{|c|c|c|c|c|c|}
  \hline
         \multicolumn{2}{|c|}{ Simulations } 	&\multicolumn{4}{ c| }{Force, $\boldsymbol{f}$} 	\\  [0.5ex] 
          \multicolumn{2}{|c|}{ }  & ~~~~$x_f$~~~~ 	& ~~~~$y_f$~~~~		      	& ~~Direction~~	& ~~~~$|\boldsymbol{f}|$~~~~ \\ [0.5ex] \hline  
         \multirow{2}{*}{~~$\mathrm{Re} =140$~~} 
  		&  ~~A01~~     		& 3$D$ 	& 0$D$ 		& $-x$    		& $10\%U$    \\ 
		&  A02     		& 3$D$ 	& 0$D$ 		& $+x$   		& $10\%U$  \\ \hline \hline

           \multirow{12}{*}{$\mathrm{Re} =150$}     
          	&    F01  		& 1$D$	& 0.25$D$ O  	& $-x$  		& $10\%U$      \\
         	&    F02  		& 1$D$	& 0.25$D$ O  	& $+x$    		& $10\%U$    \\
         	&    F03  		& 1$D$	& 0.25$D$ I  	& $-x$    		& $10\%U$ \\ 	
         	&    F04  		& 1$D$	& 0.25$D$ I  	& $+x$     		& $10\%U$  \\ \cline{2-6}
  
         	&    F05  		& 1$D$	& 0.25$D$ I  	& $-x$  		& $5\%U$   \\
         	&    F06  		& 1$D$	& 0.25$D$ I  	& $-x$    		& $20\%U$ \\
         	&    F07  		& 1$D$	& 0.25$D$ I  	& $-x$     		& $50\%U$ \\ \cline{2-6}
  
         	&    F08  		& 1$D$	& 0.50$D$ I  	& $-x$  		& $10\%U$  \\
         	&    F09  		& 1$D$	& 0.50$D$ I  	& $\sin$    	& $10\%U$  \\
         	&    F10  		& 1$D$	& 1.00$D$ I  	& $-x$    		& $10\%U$  \\ 	
         	&    F11  		& 1$D$	& 1.00$D$ I  	& $\sin$     	& $10\%U$   \\ \hline
  \end{tabular}
 \caption{Parameters of the point forces applied to the synthetic jets at $\mathrm{Re}=140$ (cases A01–A02) and $\mathrm{Re}=150$ (cases F01–F11). Forces are applied at positions $(x_f, y_f)$, where $y_f$ indicates displacement from the jet centerlines either toward the inside (I) or outside (O) of the jet region. The force magnitude is given by $|\boldsymbol{f}|$, and directions include streamwise ($+x$), counter-streamwise ($-x$), and sinusoidal forcing ($\sin$) synchronized with the piston oscillation. } 
 \label{tab:Sim_A-F}
 \end{table}

\begin{figure}[h!]
  \centering
  \includegraphics[width=0.3\textwidth]{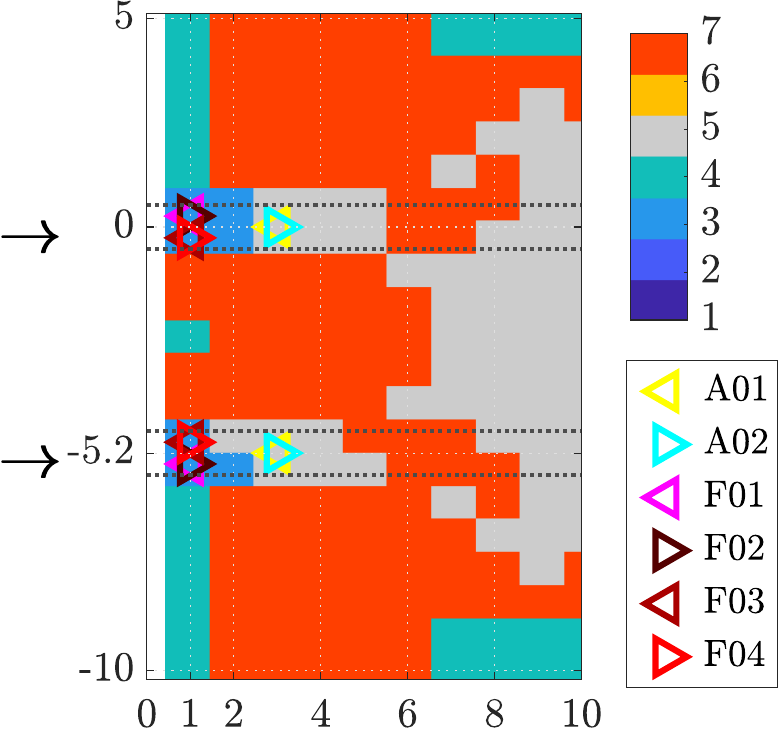} \hspace{0.2cm}
  \includegraphics[width=0.3\textwidth]{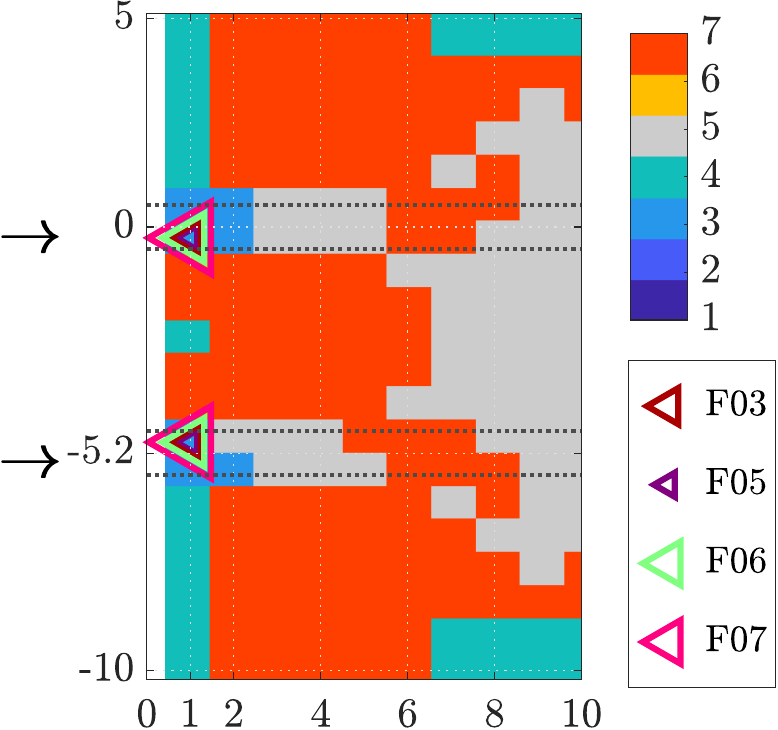}  \hspace{0.2cm}
  \includegraphics[width=0.3\textwidth]{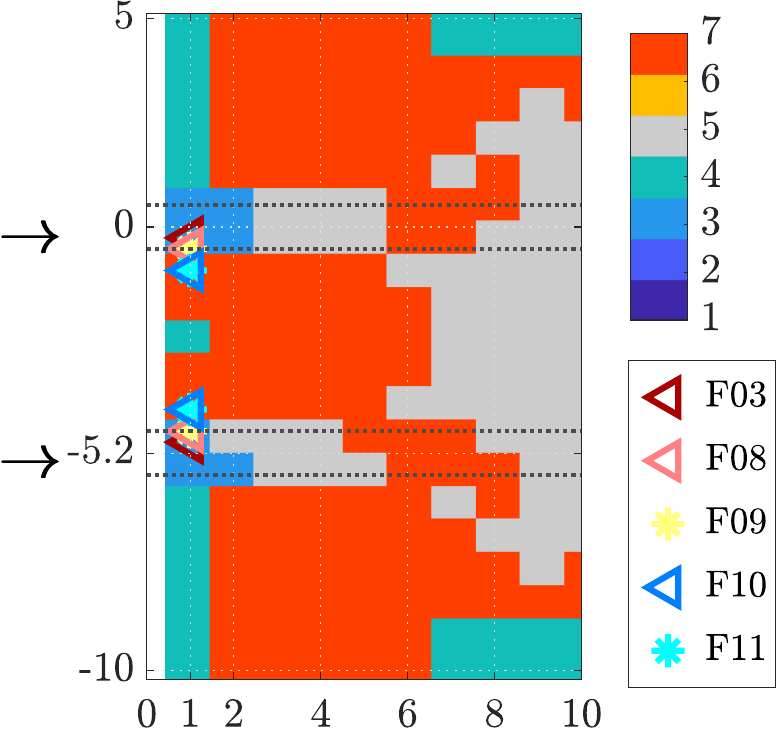}
  \caption{Scheme of position and direction of forces A01, A02, and F01-F04 over the optimal clustering solution (Fig~\ref{subfig:VQPCA_jets}) in the near region (\(x\in {[0, 10D]}\)). The dotted lines represent the wall jet positions and the arrows their centers. Symbols $\vartriangleleft$, $\vartriangleright$ and $\ast$ to represent the `$-x$', `$+x$'  and `$\sin$' force directions, respectively. Changes in symbol size correspond to changes in the magnitude of the forces. }
  \label{fig:Forces_all}
\end{figure}

The effectiveness of point forces is then analyzed based on key parameters: location, direction and magnitude, as already detailed in Table~\ref{tab:Sim_A-F}. Table~\ref{tab:Forces_all} reports the cluster where each force is applied along with the corresponding symmetry breaking cycle, cycle$_b$. Figure~\ref{fig:Forces_all} schematically represents the characteristics of the applied forces superimposed on the clustering solution. The symbols $\vartriangleleft$, $\vartriangleright$, and $\ast$ denote the force directions $-x$, $+x$, and $\sin$, respectively. The size of the symbol reflects the magnitude of the applied force.

\begin{table}[h!]
\begin{ruledtabular}
\begin{tabular}{c|ccc|cccccccccccc}
Simulations 
& \multicolumn{3}{c|}{$\mathrm{Re}=140$} 
& \multicolumn{12}{c}{$\mathrm{Re}=150$} \\
\cline{2-16}

& Original & A01 & A02 
& Original & F01 & F02 & F03 & F04 
& F05 & F06 & F07 
& F08 & F09 & F10 & F11 \\
\hline

Cluster 
& -- & 5 & 5 
& -- & 3 & 3 & 3 & 3 
& 3 & 3 & 3 
& 3 & 3 & 7 & 7 \\

$cycle_b$ 
& -- & 27 & 22 
& 22 & 25 & 26 & 29 & 28 
& 27 & 25 & 21 
& 25 & 25 & 25 & 25 \\
\end{tabular}
\end{ruledtabular}
\caption{Cluster assignment and symmetry breaking cycles for the point forces at $\mathrm{Re}$=140 (for A01-02) and $\mathrm{Re}$=150 (F01-F11).} 
  \label{tab:Forces_all}
\end{table}

\textit{Destabilizing the flow: advancing symmetry breaking ($\mathrm{Re}=140$)} \newline
The first objective is to trigger the onset of instability (symmetry breaking) in a flow that is initially symmetric but near its instability threshold, specifically at $\mathrm{Re}=140$. To this end, point forces A01 and A02 are applied within the jet streams, located in cluster $5$ (Fig.\ref{fig:VQPCA_identif}). Both forces successfully induce symmetry breaking, as indicated by their finite cycle$_b$ values (Table\ref{tab:Forces_all}). Among them, A02 is directed downstream ($+x$) and is shown to be more effective, resulting in a earlier onset of instability. These results confirm that the goal of destabilizing the flow has been achieved and reinforce that cluster 5 is a region of increased structural sensitivity.


\textit{Stabilizing the flow: delaying the symmetry breaking ($\mathrm{Re}$=150)} \newline
The second objective is to delay the onset of symmetry breaking, ideally maintaining a symmetric flow in a configuration that is initially asymmetric and close to instability ($\mathrm{Re}=150$), or at least postponing the emergence of asymmetry. Since the central point forces applied within the jet streams (cluster $5$) were shown to destabilize the flow at lower Reynolds numbers (A01 and A02), a new series of tests is conducted to evaluate whether forces applied in other regions (different clusters) and with varied orientations and magnitudes, can instead stabilize the flow.


Forces F01–F04 are positioned within the jet streams, specifically in the region corresponding to cluster $3$, slightly offset from the jet centers ($x_f=1D$, $y_f=0.25D$). These forces are applied both inside (I) and outside (O) the jet center, and in both streamwise directions ($+x$ and $-x$). \\All four configurations result in a delay in symmetry breaking compared to the original flow, as shown in Table~\ref{tab:Forces_all}. Among them, the forces applied between the jets (I), i.e., F03 and F04, are more effective in stabilizing the flow than those placed outside (F01 and F02). In particular, force F03, acting against the flow direction ($-x$), proves to be the most effective, extending the onset of symmetry breaking until cycle $29$.


The seven-cycle delay of the instability achieved by F03 effectively demonstrates the potential of the method for the second objective. However, to gain a deeper insight into the instability mechanisms and the influence of control, additional tests are conducted by varying the force parameters (position, magnitude, and temporal behavior) within the same region. Based on the promising results of F03, three directions are explored: (i) variation of the force magnitude in cases F05-F07, (ii) adjustment of the force position in F08 and F10, and (iii) introduction of a sinusoidal temporal profile in F09 and F11. As shown in Table~\ref{tab:Forces_all}, none of these alternatives surpasses F03, which remains the most effective configuration with the highest cycle$_b$.


In summary, the point forces analyzed in this section generally succeed in delaying symmetry breaking, with the exception of F07, where the high force magnitude leads to destabilization. These results suggest that the forces are strategically positioned in or near regions of structural sensitivity.\\
Considering the potential limitations of point forces (particularly in countering a flow with strong sinusoidal characteristics), an alternative strategy is also explored. This involves the use of disruptors, placed across different clusters, to control the flow and further delay the onset of symmetry breaking.


\subsubsection{Flow control with disruptors}

\begin{figure}[h!]
     	\centering
	\subfloat[Scheme of the parameters defining the position of the disruptors.] {\includegraphics[height=0.25\textheight]{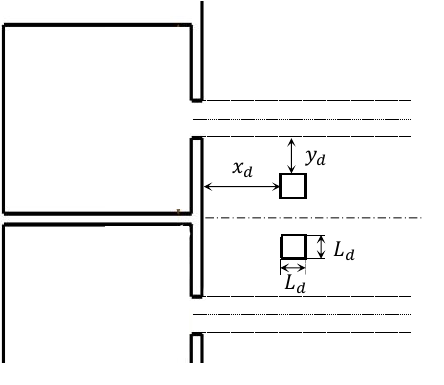}\label{subfig:Sch_d} } \hspace*{5mm}
  	\subfloat[Mesh with the disruptors] {\includegraphics[height=0.25\textheight]{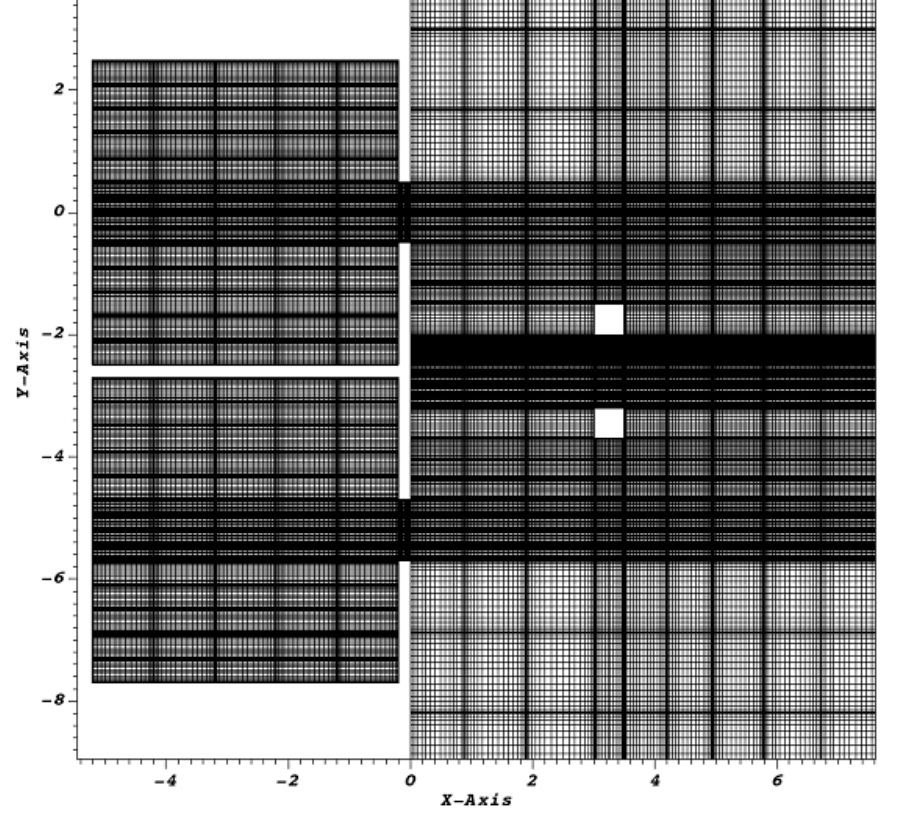}\label{subfig:msh_d} }
		\captionof{figure}{(a) Scheme of the disruptor position parameters and (b) mesh modification for D01.}
  \label{fig:Cont_D-F}
\end{figure}

As introduced in Sec.\ref{sec:Results_Cyl}, the zones identified by clustering can guide passive control strategies through geometric modifications such as square elements, referred to as disruptors. These elements are now explored in the synthetic jets configuration. Disruptors are modeled as squares with boundary conditions imposed on their edges. The width of each disruptor is defined by the parameter $L_d$, and their location is specified by $x_d$ and $y_d$: the streamwise distance from the jet exit and the normal distance from the jet walls, respectively. The mesh resolution has been increased to $1,136,592$ points, using $3508$ macroelements discretized with a polynomial order of $18$ in each direction. Figure\ref{subfig:Sch_d} illustrates the disruptor parameters, while Fig.~\ref{subfig:msh_d} shows the mesh modifications introduced to accommodate them. 

Seven disruptor configurations, labeled D01–D07, are investigated. These devices are located at $x_d \in [1D, 3D]$ and $y_d \in [0.5D, 1.5D]$, with a constant width of $L_d = 0.5D$, as summarized in Table~\ref{tab:Sim_D01-D07}. To evaluate their impact, the configuration D01 is first tested in both $\mathrm{Re}=140$ and 150 to assess whether the disruptor promotes or suppresses the instability. Based on these results, the remaining simulations are carried out at $\mathrm{Re}=150$, varying the streamwise and normal positions individually (cases D02–D05), and then in combination (cases D06–D07).


\begin{table}[h!]
\begin{ruledtabular}
\begin{tabular}{c|ccccccc}
Disruptor 
& \multicolumn{7}{c}{Simulations} \\
\cline{2-8}
Parameters 
& D01 & D02 & D03 & D04 & D05 & D06 & D07 \\
\hline
$x_d$ 
& 3$D$ & 1$D$ & 2$D$ & 3$D$ & 3$D$ & 1$D$ & 2$D$ \\

$y_d$ 
& 1$D$ & 1$D$ & 1$D$ & 0.5$D$ & 1.5$D$ & 1.5$D$ & 1.5$D$ \\

$L_d$ 
& 0.5$D$ & 0.5$D$ & 0.5$D$ & 0.5$D$ & 0.5$D$ & 0.5$D$ & 0.5$D$ \\
\end{tabular}
\end{ruledtabular}
\caption{Parameters $x_d$, $y_d$, and $L_d$ of disruptors D01--D07.}
\label{tab:Sim_D01-D07}
\end{table}

\begin{figure}[h!]
  \centering
  \includegraphics[height=0.3\textheight]{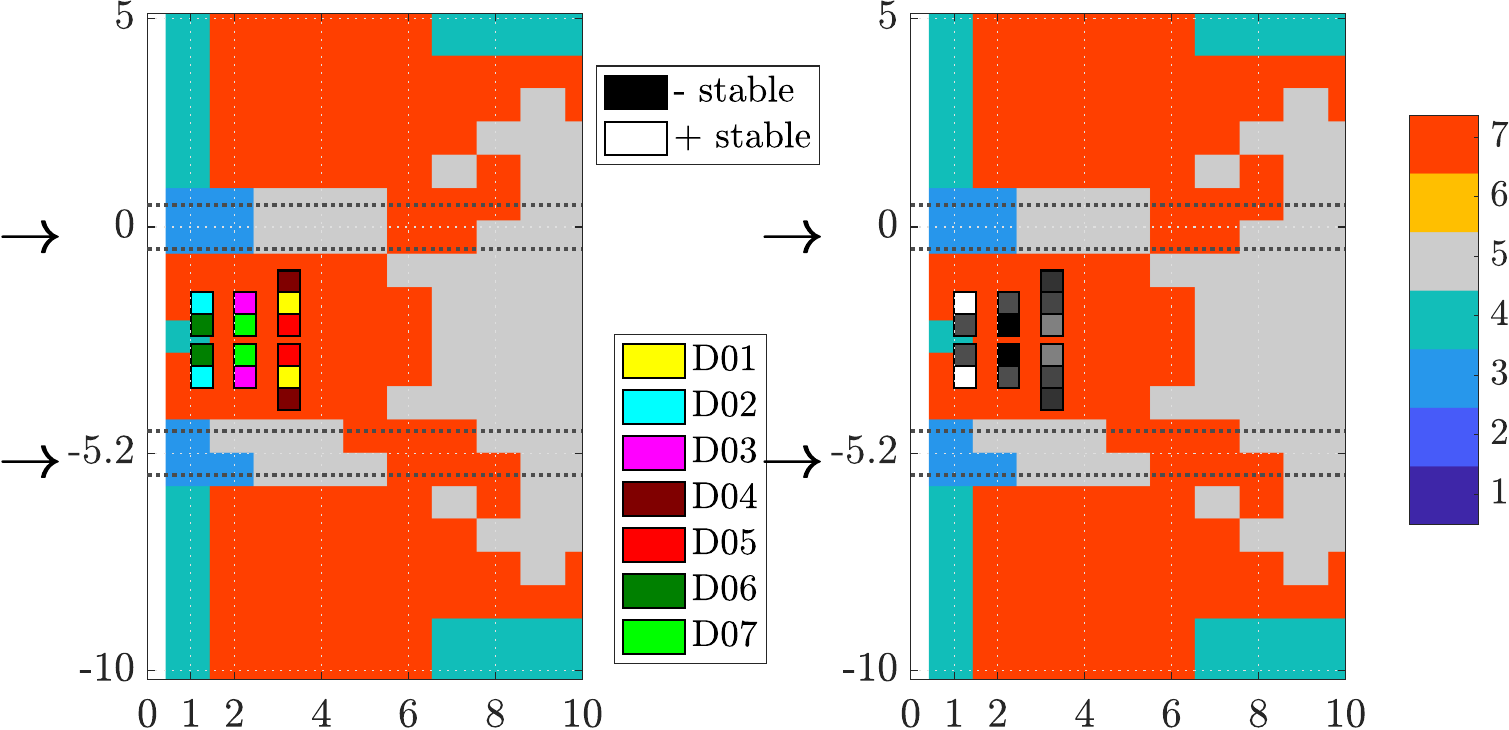}
   \caption{Scheme of the position and dimensions of the disruptors D01-D07 (left) and the stability of the simulation with them, measured by cycle$_b$, (right) over the clustering solution with $q$=1 and from 3 to 5 clusters ($k$) in the region of interest ($x \in [0,7D]$). Dotted lines represent the positions of the wall jets, and arrows indicate their centers.}
  \label{fig:Cont_E}
\end{figure}

The influence of disruptors on flow instability is now assessed, with the objective of delaying (or ideally preventing) the symmetry breakdown of synthetic jets at $\mathrm{Re}$=150. The disruptors are placed in the region where the recirculation bubbles form, a zone not previously explored in the force-based analysis. This region corresponds to cluster number $7$, located between the two jet streams, as identified by the clustering algorithm (see Fig.\ref{fig:VQPCA_identif}). The effects of varying disruptor positions are investigated for cases D01–D07 (Table\ref{tab:Sim_D01-D07}). Figure~\ref{fig:Cont_E} schematizes the positions of the disruptors superimposed on the optimal clustering solution of the reference case, with color indicating the relative stability achieved. Stability is quantified by the delay in the symmetry-breaking cycle (cycle$_b$), where a longer delay signifies increased stability. A summary of cycle$_b$ for all disruptor configurations is reported in Table~\ref{tab:Cont_150_d}.


%

\begin{table}[h!]
\begin{ruledtabular}
\begin{tabular}{c|cccccccc}
Simulation 
& Original & D01 & D02 & D03 & D04 & D05 & D06 & D07 \\
\hline
$cycle_b$ 
& 22 & 39 & 61 & 40 & 37 & 46 & 40 & 31 \\
\end{tabular}
\end{ruledtabular}
\caption{Symmetry breaking cycles, $cycle_b$, of the simulations of the synthetic jets at $\mathrm{Re}=150$ without control (Original) and with disruptors D01--D07.}
\label{tab:Cont_150_d}
\end{table}

To evaluate the impact of disruptors on flow stability, they are introduced into the synthetic jet configuration at Reynolds numbers $\mathrm{Re}=140$ and $150$, representing symmetric and asymmetric cases, respectively. The initial disruptors, labeled D01, are located near the recirculation bubbles, between the two jets in cluster number $7$, as illustrated in Figure~\ref{fig:Cont_E}.
At $\mathrm{Re}=140$, the flow retains its symmetric character with the D01 disruptors. However, at $\mathrm{Re}=150$, a significant delay in symmetry breaking is observed compared to the uncontrolled case. Asymmetry becomes apparent only at cycle $39$, $17$ cycles later than in the original configuration. This notable delay highlights the potential of disruptors to stabilize the flow, motivating further investigation into their optimal placement within cluster number $7$.


The impact of varying the disruptor position relative to the D01 configuration is now examined. First, in disruptors D02 and D03, the streamwise position $x_d$ is altered while keeping the normal position $y_d$ constant (see Table~\ref{tab:Sim_D01-D07}). Among these, the configuration with the lower $x_d$ yields a higher cycle$_b$ (Table~\ref{tab:Cont_150_d}), indicating a more stabilizing effect. Next, disruptors D04 and D05 explore variations in the $y$-position while maintaining a fixed $x_d$. In this case, greater stability is observed for higher $y_d$ values, as reflected in the corresponding cycle$_b$ values.


To explore potential synergies between the most favorable streamwise and normal positions, disruptors D06 and D07 combine the optimal $x_d$ values from D02 and D03 with the highest $y_d$ position tested ($y_d = 1.5D$). Based on previous trends, D06 (which has the smallest $x_d$ and largest $y_d$) would be expected to produce the highest stability. Surprisingly, the results indicate that D05 remains the most stabilizing case when $y_d = 1.5D$, and D02 achieves the highest overall cycle$_b$.

These results suggest that the relationship between disruptor position and flow stability is not monotonic. As shown in Fig.~\ref{fig:Cont_E}, achieving optimal placement requires a deeper physical understanding of the flow dynamics and their interaction with disruptors, beyond a simple parametric sweep.

\begin{figure}[h!]
  \centering
    \subfloat[Injection: $t=680$ and $\varphi=4\pi/5$] {\includegraphics[trim ={1.5cm 5.0cm 2cm 2.2cm},clip, width=0.7\textwidth]{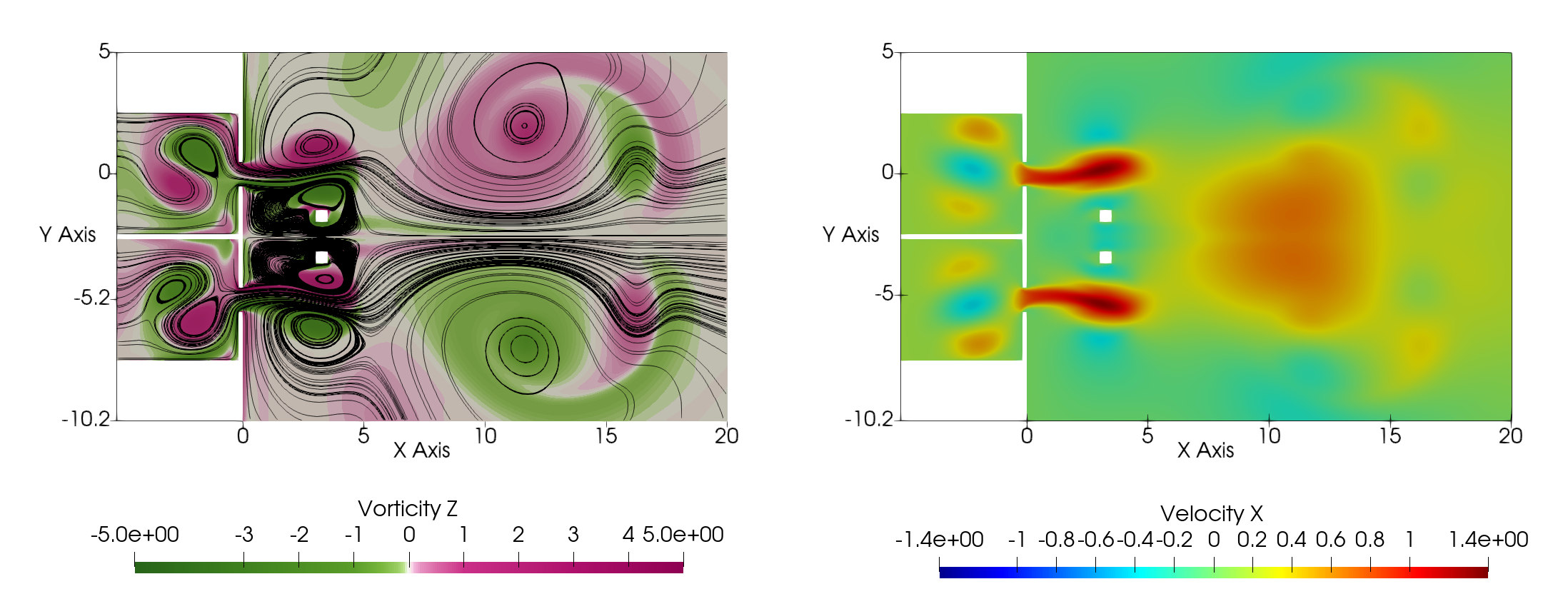}\label{subfig:Control_Re150_x3y1_Inj}}  \\
     \subfloat[Suction: $t=690$ and $\varphi=7\pi/5$] {\includegraphics[trim ={1.5cm 5.0cm 2cm 2.2cm},clip, width=0.7\textwidth]{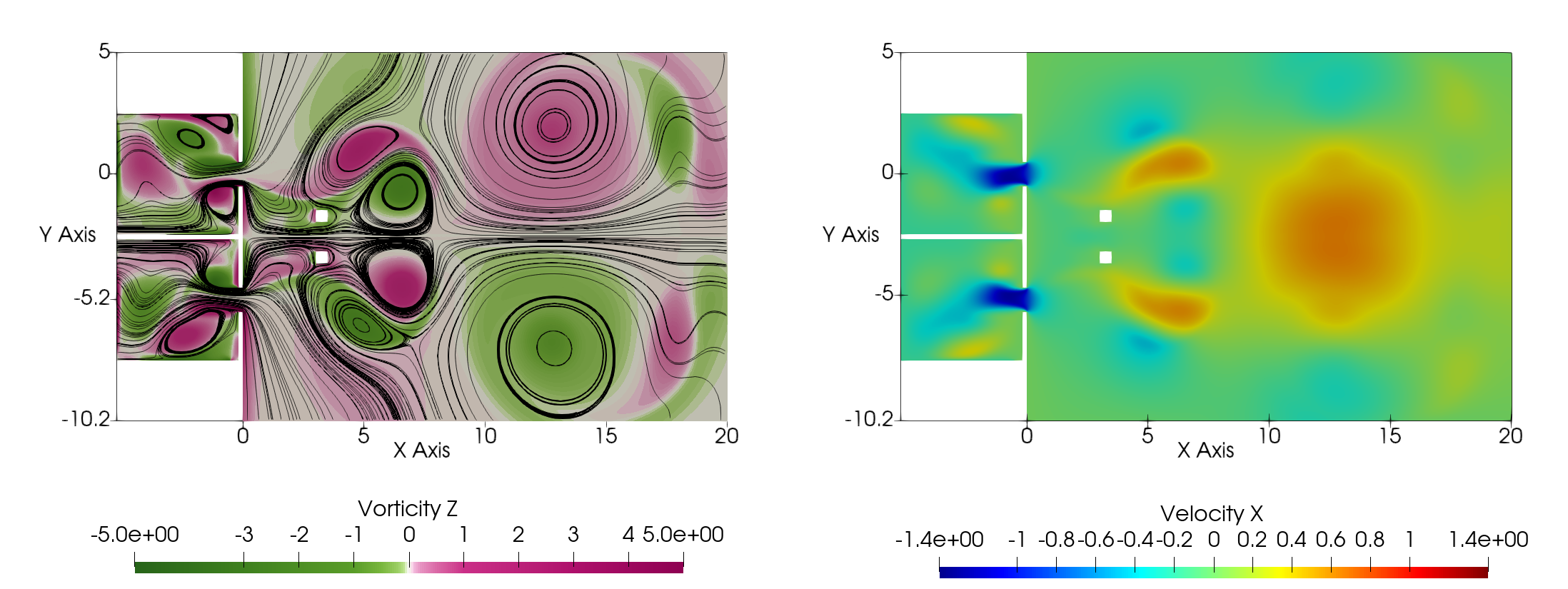}\label{subfig:Control_Re150_x3y1_Suct}} \\
      \subfloat[End of the cycle: $t=700$ and $\varphi=2\pi$] {\includegraphics[trim ={1.5cm 5.0cm 2cm 2.2cm},clip, width=0.7\textwidth]{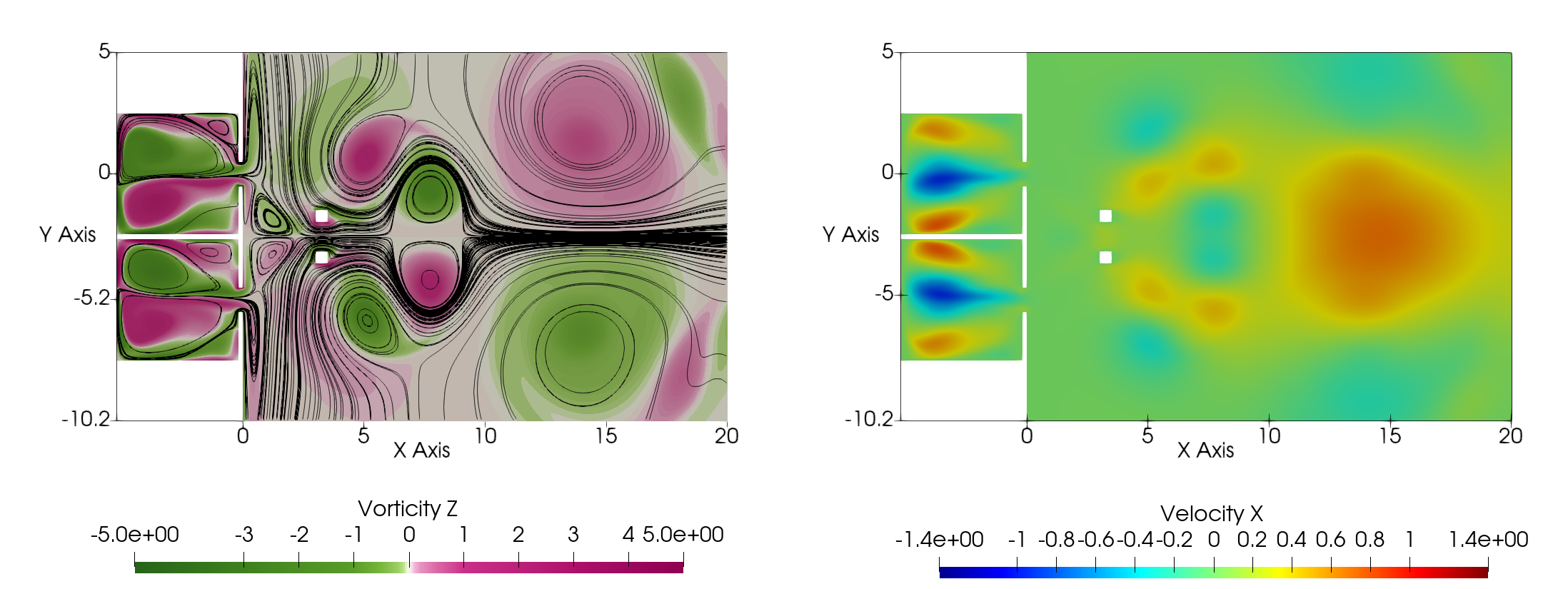}\label{subfig:Control_Re150_x3y1_End}} \\
      \includegraphics[trim ={1.5cm 0.5cm 0cm 24cm},clip, width=\textwidth]{Re150_x3y1.0069.png} \\
      \caption{Topology of synthetic jets with disruptors D01 at three snapshots. They belong respectively to the (a) injection, (b) suction, and (c) the end of the cycle of the 21st cycle (before symmetry breaking). Each subfigure displays streamlines and vorticity contours (left) and streamwise contours (right).}
  \label{fig:Control_Re150_x3y1_all}
\end{figure}

To better understand the mechanisms underlying the stabilizing effect of disruptors, the flow patterns of case D01 at $\mathrm{Re} = 150$ are analyzed in detail. Figure~\ref{fig:Control_Re150_x3y1_all} presents vorticity contours, streamlines, and streamwise velocity contours at three characteristic moments of the cycle: during injection, suction, and at the end of the cycle.

During the injection phase (Fig.\ref{subfig:Control_Re150_x3y1_Inj}), the jet streams are actively expelled from the cavity. The downstream region remains influenced by the flow structures generated during the previous cycle, leading to the merging of jet streams into a single jet. In the suction phase (Fig.\ref{subfig:Control_Re150_x3y1_Suct}), the jets move apart and their vortical structures adopt a diagonal configuration. At the end of the cycle (Fig.~\ref{subfig:Control_Re150_x3y1_End}), the flow is organized into two distinct regions: (i) between the disruptors and the jet exits, where a recirculation bubble forms, and (ii) downstream of the disruptors, where vortices continue to propagate.

Among these observed patterns, two key structures (jet streams and recirculation bubbles) were previously identified by the clustering algorithm (Fig.~\ref{fig:VQPCA_identif}), with disruptors strategically placed within the latter. This correlation suggests that disruptors influence the formation and evolution of recirculation bubbles that, in turn, play a critical role in modulating the stability of the flow.

\begin{figure}[h!]
  \centering
    \subfloat[D03: cycle$_b = 40$] {\includegraphics[trim ={1.5cm 5.0cm 50cm 2.2cm},clip, width=0.31\textwidth]{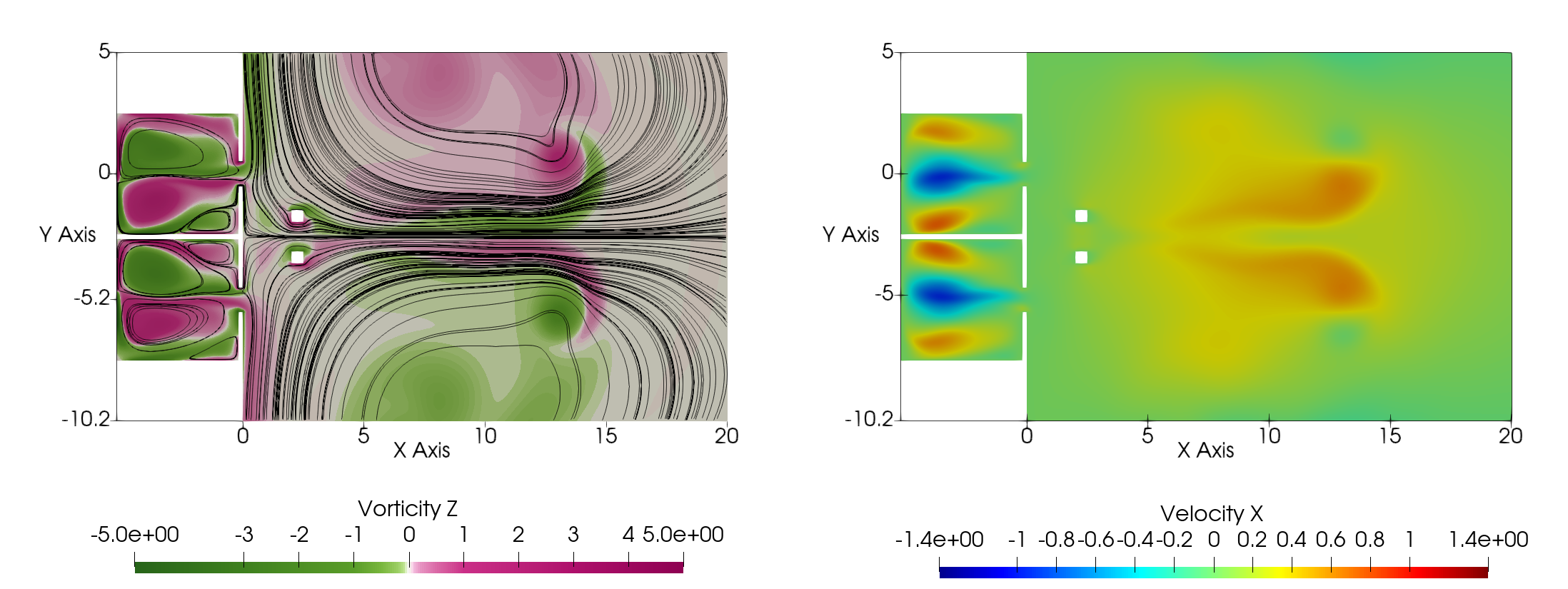}\label{subfig:Control_D01vort}}  \hspace{0.1cm}           
     \subfloat[D05: cycle$_b = 46$] {\includegraphics[trim ={1.5cm 5.0cm 50cm 2.2cm},clip, width=0.31\textwidth]{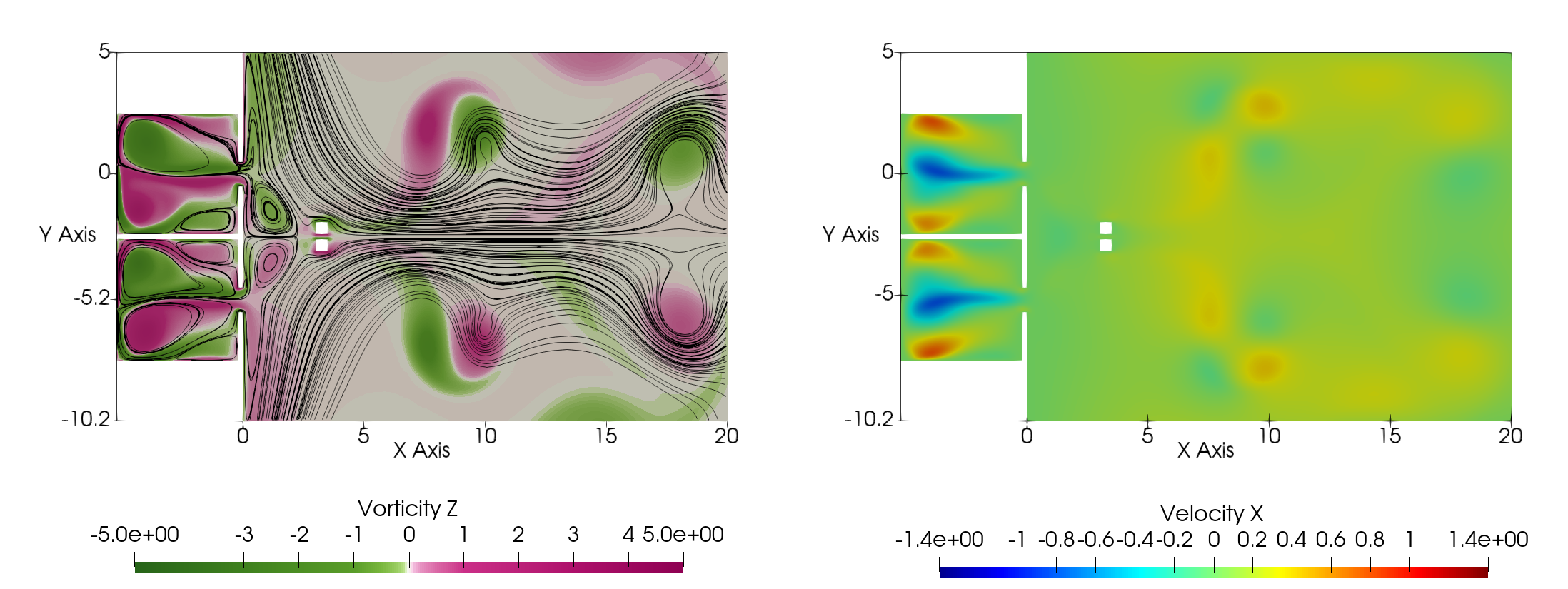}\label{subfig:Control_D05vort}}   \hspace{0.1cm} 	    
      \subfloat[D02: cycle$_b = 61$] {\includegraphics[trim ={1.5cm 5.0cm 50cm 2.2cm},clip, width=0.31\textwidth]{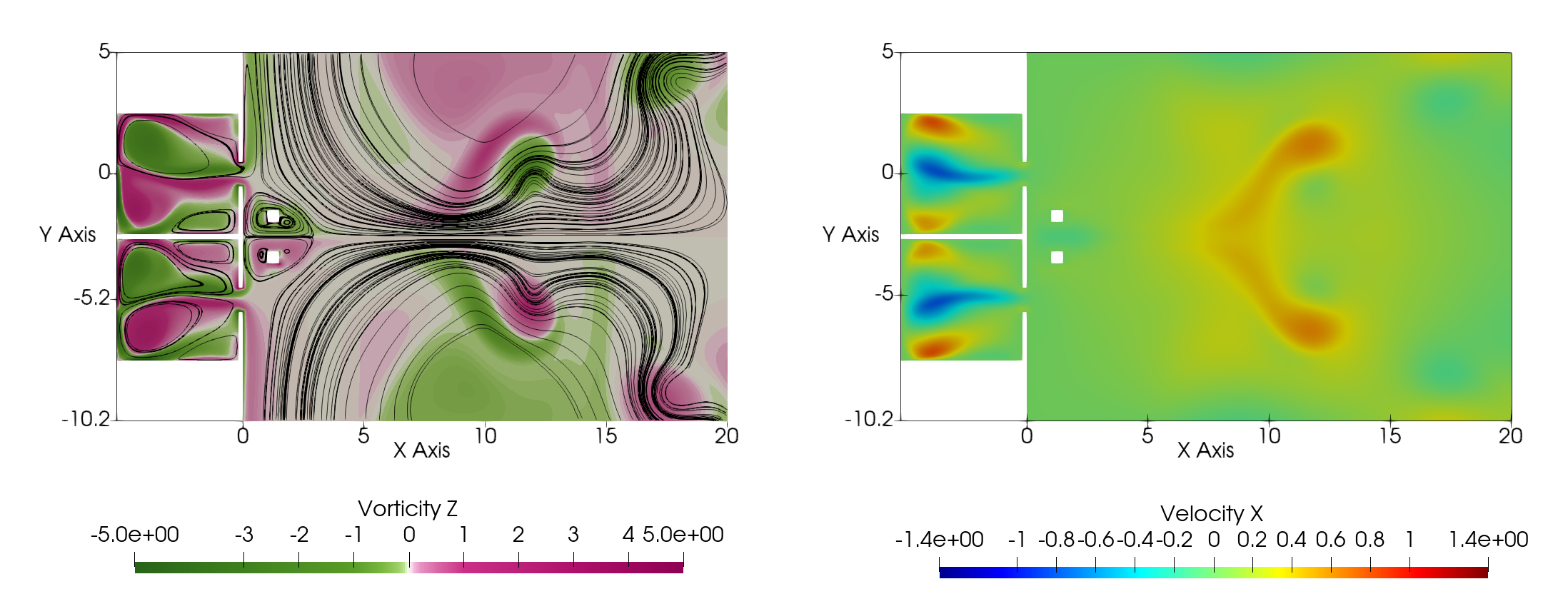}\label{subfig:Control_D02vort}} \hfill \\ 
      \includegraphics[trim ={1.5cm 0.5cm 40cm 24cm},clip, width=0.5\textwidth]{Re150_x3y1.0069.png} \\
      \caption{Streamlines and vorticity of synthetic jets at $\mathrm{Re}$=150 with disruptors D03, D05 and D02 at the end of the 21st cycle. They are order in terms of increasing stability, measured for their cycle$_b$: D03 $<$ D05 $<$ D02.}
  \label{fig:Control_bubbles}
\end{figure}

To further investigate this hypothesis (the recirculation bubble contributes to flow stabilization), Figure~\ref{fig:Control_bubbles} shows streamlines and velocity contours in the vicinity of the recirculation bubbles for cases D03, D05, and D02 (arranged in order of increasing stability). The visualizations reveal that the shape and structure of the recirculation bubble are notably influenced by the position of the disruptors. In particular, case D02, which exhibits the greatest delay in symmetry breaking, features two symmetric recirculation bubbles that closely resemble those observed in the stable configuration at $\mathrm{Re} = 140$.

As seen in Fig.~\ref{fig:Jets_vort}, these bubbles take on an oval shape, are spatially detached from jet-generated vortices, and exhibit a clear symmetry, with their shared centerline forming a significant portion of their boundary. This evidence supports the interpretation that the presence and structure of the recirculation bubbles play a stabilizing role in the flow dynamics. Specifically, by promoting the formation of these symmetric bubbles at Reynolds numbers that exceed the critical threshold ($\mathrm{Re}>140$), the disruptors contribute to postponing the onset of the symmetry-breaking instability.

The level of flow stabilization achieved in this study is notably significant compared to previous works. In the present case, disruptors, such as D02, maintained flow symmetry for up to $60$ cycles. In contrast, Rizzetta et al.\cite{Rizzetaetal1997} and Le Clainche et al.\cite{LeClainchePerez2020} considered the flow to have reached a saturated state by the $10^{\mathrm{th}}$ cycle in their simulations of two-dimensional and axisymmetric synthetic jets, respectively. Palomo et al.\cite{Palomoetal2019} adopted a more conservative threshold, defining the onset of the saturated regime at the $15^{\mathrm{th}}$ cycle for concentric synthetic jets. Furthermore, Hayes-McCoy et al.\cite{Hayesetal2008} reported no further development in the flow beyond the $5^{\mathrm{th}}$ cycle for an axisymmetric synthetic jet at $\mathrm{Re}=500$ and $\mathrm{St}=0.08$. These references suggest that symmetry breaking tends to emerge during the saturated regime, typically assumed to occur around the $10^{\mathrm{th}}$ cycle. As such, it is reasonable to hypothesize that the instability observed in our simulations corresponds to a physically relevant mode rather than being solely the result of numerical artifacts.


In summary, flow control was successfully achieved for two cases near the instability threshold. In the first case, an initially symmetric flow was destabilized, while in the second case an unstable flow was stabilized. Point forces were used to target cluster $5$ (jet stream region) for destabilization, whereas stabilization was achieved by applying point forces in cluster $3$ and, most effectively, disruptors in cluster $7$ (recirculation bubble region). These control outcomes were made possible by the ability of the clustering algorithm to identify flow structures that are dynamically and structurally significant.


These results highlight the effectiveness of clustering a framework for analyzing the dynamics of complex fluid flows. Unsupervised identification of coherent flow structures, such as jet streams and recirculation bubbles, proved essential to identify potential regions where control measures can be applied. By targeting specific clusters, this study was able to successfully delay or advance the onset of symmetry breaking. This suggests that clustering not only captures dynamically relevant flow features, but also provides a promising framework for directing and optimizing control strategies in complex fluid systems.

\end{document}